\documentclass[twocolumn,showpacs,preprintnumbers,prb]{revtex4-1}
\usepackage{amssymb}
\usepackage[dvips]{color}
\usepackage{epsfig}
\usepackage{dcolumn}
\usepackage{bm}

\begin{document}

\title{Non-Zeeman splitting for a spin-resolved STM with a Kondo adatom in a spin-polarized two-dimensional electron gas}

\author{A. C. Seridonio$^1$, F. S. Orahcio$^2$, F. M. Souza$^3$, and M. S. Figueira$^2$}
\affiliation{$^1$Departamento de F\'isica e Qu\'imica, Universidade Estadual Paulista ``J\'ulio de Mesquita Filho", 15385-000, Ilha Solteira, SP, Brazil\\
$^2$Instituto de F\'{i}sica, Universidade Federal Fluminense, 24210-340 Niter\'{o}i, RJ, Brazil\\
$^3$Instituto de F\'isica, Universidade Federal de Uberl\^andia, 38400-902, Uberl\^{a}ndia, MG, Brazil.}

\begin{abstract}
We theoretically investigate the spin-resolved local density of states (SR-LDOS) of a spin-polarized two-dimensional electron gas in the presence of a
Kondo adatom and a STM probe. Using Green function formalism and the atomic approach in the limit of infinite Coulomb correlation,
it is found an analytical SR-LDOS expression in the low temperature regime of the system. This formal result is given in terms of phase shifts originated by the adatom scattering and Fano interference.
The SR-LDOS is investigated as a function of the probe position and different Fano factors.
Our findings provide an alternative way to spin-split the Kondo resonance without the use of huge magnetic
fields, typically necessary in adatom systems characterized by large Kondo temperatures.
We observe a non-Zeeman spin-splitting of the Kondo resonance in the total LDOS, with one spin-component pinned
around the host Fermi level. Interestingly, this result is in accordance to recent experimental data reported in Phys. Rev. B 82, 020406(R) (2010).
\end{abstract}


\date[Date: ]{\today}
\maketitle

\section{Introduction}
\label{sec1}

The scattering of electrons by a magnetic impurity in a metallic environment
is responsible for the manifestation of the Kondo effect.\cite{Hewson} This phenomenon occurs as a result of an antiferromagnetic
coupling between the localized spin at the impurity and the surrounding conduction
electrons of the host. In particular, at temperatures much lower than the Kondo temperature $T_{K}$, an electron cloud emerges
to screen the magnetic moment placed at the impurity site. Thus, a sharp resonance
pinned at the Fermi energy appears in the impurity density of states and characterizes the formation of the Kondo peak.
Such effect was first observed in resistivity measurements of magnetic alloys, later
in transport properties of quantum dots (QDs) performed in a two-dimensional electron gas
(2DEG).\cite{QD1,QD2,QD3,QD5,QD6,QD7,QD10,QD11,QD12,QD13} More recently, Kondo effect was also measured using scanning tunneling microscope (STM)
in the presence of impurities deposited on metallic surfaces.\cite{STM1,STM2,STM3,STM4,STM5,STM6,STM7,STM8,STM9,STM10,STM11,STM12,STM13,STM14,STM15,STM16}

In the context of unpolarized scanning tunneling microscope (STM) probes, the conductance exhibits the Fano line shape
due to the quantum interference between the transport channels given by the conduction bands and the adatom.\cite{STM13,STM14,STM15,STM16,Fano1,Fano2}
Notably, for STM probes not very close to the metallic host and blue in the low temperature regime,
the STM device probes the local density of states (LDOS) of the sample.

In the case of spin-polarized STM probes, interesting new features emerge as the spin-splitting of the Fano-Kondo profile
of the conductance and the Fano-Kondo spin-filter.\cite{SPSTM1,SPSTM2,SPSTM3,SPSTM4} In this scenario, several experimental and theoretical
works discuss related phenomena employing ferromagnetic leads coupled to QDs and adatoms \cite{SPSTM5,FM1,Yunong,FM2,FM3,FM4,FM5,FM6,FM7,FM8,FM88,FM9,FM10,FM11,FM12,Kawahara2010,Nygard2004,new1,new2,new3,new4,new5}.
In particular, in the emerging field of spintronics, the interplay between the Fano-Kondo effect and the ferromagnetism of a metallic environment, plays a crucial role in the manufacturing of novel spintronics devices.

\begin{figure}[tbh]
\includegraphics[clip,width=0.44\textwidth]{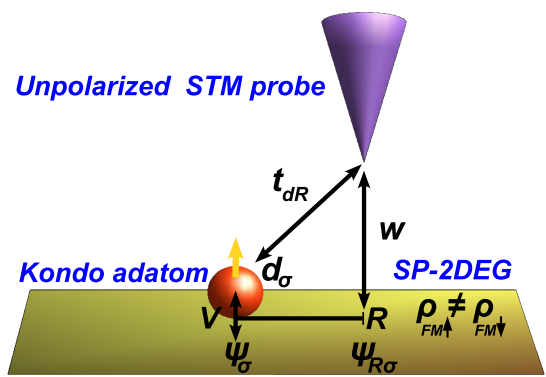}
\caption{\label{fig:Pic1}(Color online) STM device composed by an unpolarized probe and a Kondo adatom hybridized to a SP-2DEG.
The parameters $V$, $t_{dR}$ and $w$ are hopping terms. The letters $d_{\sigma}$, $\Psi_{\sigma}$ and $\Psi_{R\sigma}$ denote, respectively,
the fermionic annihilation operators to the adatom, to the site of the host side-coupled to the adatom and to the site just below the probe.}
\end{figure}

In this work, we report an analytical description of the spin-resolved local density of states (SR-LDOS), for an unpolarized STM probe,
hybridized with a single Kondo adatom in a ferromagnetic (FM) island,  considered here as a spin-polarized two-dimensional electron gas (SP-2DEG).
Such analysis is done in the framework of the single impurity Anderson model (SIAM) \cite{SIAM}, using the atomic approach with infinite Coulomb
correlation \cite{AP1,AP2,Ufinito} in order to determine the adatom Green function (GF).

The main result of our simulations is the emergence of an asymmetric spin-splitting in the Kondo peak,
which agrees with some experimental works.\cite{SPSTM5,FM9,FM11,Kawahara2010,Nygard2004}
Here we highlight the experiment done by S. L. Kawahara \textit{et al} \cite{Kawahara2010},
where the measured LDOS shows that one spin channel has a Kondo peak pinned at the vicinity of the host Fermi level,
while the opposite is shifted by the spin-polarization of the 2DEG forming the island surface.
We note that the behavior we see here is not similar to the usual Zeeman
splitting due to an external magnetic field.\cite{QD1,QD2,QD6,QD7,STM12} In the present model there is no
external fields.

Contrary to the QDs systems, where the Kondo temperature is of the order of milliKelvin and the temperature could be
tuned to observe the suppression of the Kondo resonance, in most STM systems,\cite{comment1} the Kondo temperature is of the order of
tens of Kelvin and the experimental STM setups do not allow the temperature variation in such range to observe the evolution
of the Kondo effect. On the other hand, to verify the splitting of the Kondo peak in STM experiments, it is necessary to
apply a magnetic field of hundreds of Tesla, which is not feasible. Thus, the type of experiment proposed in our work could
be useful to produce a splitting of the Kondo resonance without huge magnetic fields.

This paper is organized as follows. In Sec. \ref{sec2}, we show the theoretical model for the STM in terms of the Anderson Hamiltonian.
We derive in Sec. \ref{sec3}, the SR-LDOS formula with and without a STM probe. For both cases in
the low temperature limit, we show that the SR-LDOS expression can be labeled by phase shifts due to the adatom
scattering and Fano interference. In Sec. \ref{sec4}, we discuss the results of the host Fano parameter, which displays spin-polarized
Friedel oscillations \cite{Friedel1,GFM,Friedel2} and we present a wide analysis of the SR-LDOS as a function of the bias-voltage,
for different probe positions and Fano factors. The atomic approach is employed considering an infinite Coulomb correlation in
order to calculate the adatom GF. We apply our formulation in Sec. \ref{sec5} to describe the Kondo peak splitting found in the
experiment of Ref. [\onlinecite{Kawahara2010}]. The conclusions appear in Sec. \ref{sec6} and in the Appendices,
we give details of the derivations of the atomic GF for the adatom as well as for the host Fano factor.

\section{Theoretical Model}
\label{sec2}

\subsection{Hamiltonian}

In Fig. \ref{fig:Pic1}, we represent an unpolarized STM probe coupled to the FM island hybridized to the Kondo adatom
deposited on its surface. Note that when the hopping term $t_{dR}>>w$, the setup behaves as a Single Electron Transistor (SET) \cite{QD1,QD2},
which we call \emph{peak limit}, due to the emergence of the Kondo resonance in the LDOS energy profile as we shall see. The other limit we call \emph{dip limit},
it comprises the cases  $t_{dR}<<w$ and $t_{dR}\simeq w$, which resemble the T-Shaped QD device \cite{QD7} characterized by a Fano-Kondo dip.

The system we investigated is described according to the Hamiltonian
\begin{equation}
\mathcal{H}=\mathcal{H}_{FM}+\mathcal{H}_{probe}+\sum_{\sigma}h_{tun}^{\sigma}. \label{eq:1}\end{equation}
The first term represents the SIAM \cite{SIAM}
\begin{eqnarray}
\mathcal{H}_{FM} & = & \sum_{\vec{k}\sigma}\varepsilon_{k\sigma}c_{\vec{k}\sigma}^{\dagger}c_{\vec{k}\sigma}+\sum_{\sigma}E_{d}d_{\sigma}^{\dagger}d_{\sigma}+\sum_{\vec{k}\sigma}V_{dk\sigma}(c_{\vec{k}\sigma}^{\dagger}d_{\sigma}\nonumber \\
 & + & d_{\sigma}^{\dagger}c_{\vec{k}\sigma})+Ud_{\uparrow}^{\dagger}d_{\uparrow}d_{\downarrow}^{\dagger}d_{\downarrow} ,\label{eq:2}\end{eqnarray}
that assumes the island as a SP-2DEG described by the operator $c_{\vec{k}\sigma}^{\dagger}$ ($c_{\vec{k}\sigma}$) for the creation (annihilation) of an electron in a quantum state labeled by the wave vector $\vec{k}$, spin $\sigma$ and energy
\begin{equation}
\varepsilon_{k\sigma}=D_{\sigma}k_{F\sigma}^{-1}\left(k-k_{F\sigma}\right) , \label{eq:energy}
\end{equation}
that depends on the spin-polarized band half-widths $D_{\sigma}$ forming the host Fermi sea and the wave numbers $k_{F\sigma}$ evaluated at the Fermi energy $\varepsilon_{F}\equiv0$.

For the adatom, $d_{\sigma}^{\dagger}$ ($d_{\sigma}$) creates (annihilates) one electron with spin $\sigma$ in state $E_d$. The third term hybridizes the adatom level
and the host continuum of states. The coupling matrix element is modeled as
\begin{equation}
V_{dk\sigma}=\frac{V}{\sqrt{N_{FM\sigma}}}\frac{\Gamma^{2}}{\Gamma^{2}+\varepsilon_{k\sigma}^{2}} ,
\label{eq:L}\end{equation}
which obeys a Lorentzian behavior for a sake of simplicity to mimic a nonlocal coupling between the adatom and the island.
$N_{FM\sigma}$ is the number of conduction states for a given spin and $\Gamma$ is the width of this interaction around $\varepsilon_{F}$.
Coulomb correlation between two electrons with opposite spins at the adatom site is also taken into account and is represented by the letter $U$.
Here we assume this parameter as infinite in order to implement the atomic approach \cite{AP1,AP2} that gives the adatom GF. In particular,
taking the limit $\Gamma\gg\varepsilon_{k\sigma}$ in Eq. (\ref{eq:L}), we obtain a constant $V_{dk\sigma}$, i.e., $V_{dk\sigma}=\frac{V}{\sqrt{N_{FM\sigma}}}$.
This corresponds to the case of a site of the FM island side-coupled to the adatom, which we designate local coupling.
We mention that Eqs. (\ref{eq:energy}) and (\ref{eq:L}) were previously applied in the context of unpolarized bulk electrons in a system described by the
Kondo model.\cite{GFM} In this work, we employ the Lorentzian shape to emulate the nonlocality of the adatom-island coupling.

The FM island is considered a spin-polarized electron bath, with polarization given by
\begin{equation}
P=\frac{\rho_{FM\uparrow}-\rho_{FM\downarrow}}{\rho_{FM\uparrow}+\rho_{FM\downarrow}}.\label{eq:SP}
\end{equation}
where
\begin{equation}
 \rho_{FM\sigma}=\frac{1}{2D_{\sigma}},\label{eq:rho_spin}
\end{equation}
is the host density of states in the flat band approximation.
This quantity is expressed in terms of
\begin{equation}
D_{\sigma}=\frac{D_{0}}{\left(1+\sigma P\right)},\label{den_spin}
\end{equation}
for a given spin $\sigma$ and the unpolarized density
\begin{equation}
\rho_{0}=\frac{1}{2D_{0}} , \label{eq:den}\end{equation}
written in terms of the width $D_{0}$. The unpolarized part of the system is the conduction band given by the Hamiltonian
\begin{equation}
\mathcal{H}_{probe}=\sum_{\vec{p}\sigma}\varepsilon_{p}b_{\vec{p}\sigma}^{\dagger}b_{\vec{p}\sigma} , \label{eq:3}\end{equation}
which corresponds to the second term in Eq. (\ref{eq:1}) for free electrons in the STM probe.
Such conduction electrons are ruled by fermionic operators $b_{\vec{p}\sigma}^{\dagger}$ and $b_{\vec{p}\sigma}$.
To perform the coupling between Eqs. (\ref{eq:2}) and (\ref{eq:3}), we have to define
\begin{equation}
h_{tun}^{\sigma}=\sum_{\vec{p}}b_{\vec{p}\sigma}^{\dagger}\left(w \Psi_{R\sigma}+t_{dR} d_{\sigma}+H.c.\right) , \label{eq:4}\end{equation}
as the spin tunneling Hamiltonian that hybridizes the STM probe conduction states with those in the island and the adatom.
The former hopping parameter, we consider proportional to the energy independent term $w$, with the fermionic operator
\begin{equation}
\Psi_{R\sigma}=\frac{1}{\sqrt{N_{FM\sigma}}}\sum_{\vec{k}}e^{i\vec{k}.\vec{R}}c_{\vec{k}\sigma}, \label{eq:PSI_R}
\end{equation}
which describes a conduction state at the site $\vec{R}$ laterally displaced from the adatom.
This operator admits an expansion in plane waves, due to the assumption of an infinite 2DEG forming the island surface. The second hybridization parameter in Eq. (\ref{eq:4})
is proportional to the adatom operator $d_{\sigma}$ and the spatial dependent hopping
\begin{equation}
t_{dR}=t_{do}\exp(-k_{F}R) , \label{eq:td}\end{equation} which provides a decreasing STM probe-adatom coupling. This characteristic ensures a vanishing behavior for huge lateral displacements, which was already employed in the literature \cite{STM3,SPSTM2,SPSTM3}.

\subsection{The atomic approach}
\label{secAP}
In order to implement the atomic approach \cite{AP1,AP2,Ufinito}, we consider the adatom-island coupling as local. Thus, we begin
\begin{eqnarray}
\mathcal{H}_{FM} & =\sum_{\vec{k}\sigma}\varepsilon_{k\sigma}c_{\vec{k}\sigma}^{\dagger}c_{\vec{k}\sigma}+\sum_{\sigma}E_{d}X_{d,\sigma\sigma}\nonumber\\
 & +V\left(X_{d,0\sigma}^{\dagger}\Psi_{\sigma}+\Psi_{\sigma}^{\dagger}X_{d,0\sigma}\right),\label{SIAM2}
\end{eqnarray}
which is derived from from Eq. (\ref{eq:2}) taking into account $V_{dk\sigma}=\frac{V}{\sqrt{N_{FM\sigma}}}$. Here $X$ is the Hubbard operator\cite{Hubbard1,Hubbard2}
that project out the doubly occupied state from the adatom to ensure the limit of infinite Coulomb correlation, and $\Psi_{\sigma}$ is $\Psi_{R\sigma}$ [Eq. (\ref{eq:PSI_R})] evaluated at the origin.
This Hamiltonian is useful for the calculation of the adatom GF and consequently the system SR-LDOS.

This GF is based on an extension of the Hubbard cumulant expansion also applicable to the Anderson lattice with impurity-host couplings
treated as perturbations.\cite{Hubbard2} The use of this expansion allows to express the exact GF in terms of an unknown effective cumulant. In a previous work, we have studied the Anderson lattice with an approximate effective cumulant obtained from the atomic limit of the
model in a procedure that we call as the zero band width (ZBW) approximation. The advantage is that such method includes all the higher
order cumulants absent in our previous diagrammatic calculations.\cite{Hubbard2}

The method presents some similarities to the exact diagonalization (ED) for the SIAM. The ED is a brute-force method to solve the Hamiltonian
treated as a discrete bath by considering the impurity coupled to a finite number of conduction sites ($N_c$) of the host band. In principle,
it is an exact method as the name implies, but its limitation relies in the number of conduction sites considered, the Hilbert space grows extremely
fast when $N_c$ is enlarged. The cumulant atomic approach is not equivalent to the ED, but employs an exact diagonalization on a reduced bath with
two sites for the Anderson Hamiltonian, one is the impurity site and the other is a representative one for the conduction band.
They compose the starting point of the method, the ZBW approximation. The representative site of the band will be determined by using the Friedel sum rule as we shall see. Employing the Lehmann representation, the adatom atomic GF is used as well as the approximated cumulants,
which will be considered to calculate the full GF. This GF does not present that spurious oscillatory behavior in the Kondo peak as found by the standard ED method. In the ED context, this artifact is solved by the \emph{self-energy trick}.\cite{new5}.
We also would like to mention, that the atomic approach could be easily generalized to multi-orbital Anderson model, because once the atomic solution is known, the method follows the same steps realized in the spin $S=1/2$ SIAM case.

To obtain the exact GF of an Anderson impurity (adatom), we can employ the chain approximation,\cite{Hubbard2} but considering all the possible cumulants in the expansion for the Anderson lattice. Similarly to the Feynman diagrams, it is possible to rearrange all those that contribute to the exact adatom GF by defining an effective cumulant,
determined by all the diagrams that cannot be separated by cutting a single edge ({}``proper\textquotedblright{}\ or {}``irreducible\textquotedblright{}\ diagrams).

As we are interested in the exact GF for the adatom, we use the standard definition
\begin{equation}
\mathcal{G}_{\sigma}^{dd}\left(\tau\right)=-\frac{i}{\hbar}\theta\left(\tau\right)Tr\left\{ \varrho_{FM}\left[d_{\sigma}\left(\tau\right),d_{\sigma}^{\dagger}\left(0\right)\right]_{+}\right\}\label{eq:G_d_d}
\end{equation}
in time coordinate, where $\hbar$ is the Planck constant divided by $2\pi$, $\theta\left(\tau\right)$ the step function at the instant $\tau$,
$Tr$ the trace over the eigenstates of the Hamiltonian in Eq. (\ref{eq:2}), $\varrho_{FM}$ the density matrix of the FM island and $\left[,\right]_{+}$ is the anticommutator between the adatom operators at different times.

The time Fourier transformation of Eq. (\ref{eq:G_d_d}) thus provides the adatom GF in energy coordinate,
which is then obtained by replacing the bare cumulant for the effective one calculated by following the
atomic approach with the Hamiltonian in Eq. (\ref{SIAM2}). As a result, we have
\begin{equation}
\mathcal{G}_{\sigma}^{dd}(\omega)=\ \frac{\mathcal{M}_{\sigma}^{eff}(\omega)}{1-\mathcal{M}_{\sigma}^{eff}(\omega)\mid V\mid^{2}\sum_{\vec{k}}\mathcal{G}_{\sigma}^{c}(\vec{k},\omega)\ }, \label{Eq.6}
\end{equation}
as the adatom GF in terms of the effective cumulant $\mathcal{M}_{\sigma}^{eff}(\omega)$ and the free-electron GF
\begin{equation}
\mathcal{G}_{\sigma}^{c}(\vec{k}{,}\omega)=\frac{1}{\omega-\varepsilon_{\vec{k}\sigma}+i \eta},\label{eq:AA_freeGF}
\end{equation}
where $\eta \to 0^+$. The atomic version of Eq. (\ref{Eq.6}) is given by (see Appendix A)
\begin{equation}
\mathcal{G}_{at,\sigma}^{dd}(\omega)=\frac{\mathcal{M}_{\sigma}^{at}(\omega)}{1-\mathcal{M}_{\sigma}^{at}(\omega)\mid V\mid^{2}\mathcal{G}_{\sigma}^{ZBW}(\omega)\ } , \label{Eq.6a}
\end{equation}
which results in
\begin{equation}
\mathcal{M}_{\sigma}^{at}(\omega)=\frac{\mathcal{G}_{at,\sigma}^{dd}(\omega)}{1+\mathcal{G}_{at,\sigma}^{dd}(\omega)\mid V\mid^{2}\mathcal{G}_{\sigma}^{ZBW}(\omega)} , \label{Eq.7}
\end{equation}
for the effective cumulant determined from the adatom GF calculated in Appendix \ref{ApA}, both dependent on
\begin{equation}
\mathcal{G}_{\sigma}^{ZBW}(\omega)=\frac{1}{\omega-(\varepsilon_{0\sigma}-\mu)+i \eta}, \label{Eq3.144}
\end{equation}
for an electron state, in the ZBW approximation with $\mu$ to denote the FM island chemical potential.
As we can see, Eq. (\ref{Eq3.144}) replaces all energy contributions of the original Fermi sea by two spin
dependent atomic levels, i.e., we perform the substitution
$\sum_{\vec{k}\sigma}\varepsilon_{k\sigma}c_{\vec{k}\sigma}^{\dagger}c_{\vec{k}\sigma}\rightarrow\sum_{\sigma}\varepsilon_{0\sigma}c_{0\sigma}^{\dagger}c_{0\sigma}$
in Eq. (\ref{SIAM2}) with $\varepsilon_{k\sigma}=\varepsilon _{0\sigma}$ representing the band atomic level for a given spin $\sigma$.
As this procedure overestimates the coupling of the spin-polarized conduction bands of the island with the adatom due to the concentration
of the bands at atomic levels, we have to moderate this effect, \cite{moderation} performing the substitution of $V^{2}$ by $\Delta ^{2}$
in Eqs. (\ref{Eq.6a}) and (\ref{Eq.7}), where $\Delta=\pi V^{2}\rho_{0}$ is the Anderson parameter.

To determine the adatom GF, we use the atomic cumulant $\mathcal{M}_{\sigma}^{at}(\omega)$ as effective in Eq. (\ref{Eq.6}) and verify that
\begin{equation}
\mathcal{G}_{\sigma}^{dd}(\omega)=\frac{\mathcal{M}_{\sigma}^{at}(\omega)}{1-\mathcal{M}_{\sigma}^{at}(\omega)\frac{\left\vert V\right\vert ^{2}}{2D_{\sigma}}ln\left(\frac{\omega+D_{\sigma}+\mu}{\omega-D_{\sigma}+\mu}\right)}\label{Gff_ap}
\end{equation}
provides an analytical expression in the flat band approximation. We mention that $\mathcal{M}_{\sigma}^{at}(\omega)$ is a simplification,
but it contains all the diagrams that should be presented in such way that the correspondent GF displays realistic features.

As the final step of the atomic approach implementation, we have to find adequate values of the atomic levels $\varepsilon_{0\sigma}$
that well describe the ZBW GFs in Eq. (\ref{Eq3.144}) and consequently the adatom GF. To that end, we use the condition that,
in metallic systems, the most important region for conduction electrons is placed at the Fermi energy $\varepsilon_{F}$ and
that the coupling to an Anderson impurity leads to the Friedel's sum rule \cite{Langreth}
\begin{equation}
\rho_{d,\sigma}(\varepsilon_{F})=-\frac{1}{\pi}\Im\left\{\mathcal{G}_{\sigma}^{dd}(\varepsilon_{F})\right\}=\frac{sin^{2}\left(\delta_\sigma(\varepsilon_{F})\right)}{\pi\Delta_{\sigma}} \label{fried}
\end{equation}
for the adatom spectral density. Here $\delta_\sigma(\varepsilon_{F})=\pi n_{d,\sigma}$ is the conduction phase shift at the Fermi level, $\Im$ represents the imaginary part, $\Delta_{\sigma}=\Delta(1+\sigma P)$ is the spin dependent Anderson parameter
and $n_{d,\sigma}$ is the adatom occupation with spin $\sigma$. Thus we find the atomic levels $\varepsilon_{0\sigma}$ calculating self-consistently  Eq. (\ref{fried}) together with
\begin{equation}
n_{d,\sigma}=<X_{d,\sigma\sigma}>=-\frac{1}{\pi }\int_{-\infty }^{+\infty }d\omega \Im\left\{\mathcal{G}_{\sigma}^{dd}(\omega)\right\}n_{F}(\omega).\label{G00}
\end{equation}
In Eq. (\ref{G00}), $n_{F}(\omega)$ is the Fermi-Dirac distribution.

It is important to emphasize here that the choice of the atomic approach to calculate the adatom GF is only due to its computational
simplicity and ability to obtain the Kondo peak, but we must take into account that the method presents some limitations that were extensively
discussed in the original papers \cite{AP1,AP2,Ufinito}. The SR-LDOS formulas obtained in Sec. \ref{sec3} are general,
we could employ others more powerful and precise techniques to calculate the GF of the Anderson impurity, like the Numerical
Renormalization Group (NRG) \cite{QD12,QD13,Bulla} without any modifications in the formalism.

\section{SR-LDOS}
\label{sec3}
\subsection{SR-LDOS in the scheme of phase shifts for the FM island}
\label{sec3A}

In this section we derive at the temperature range $T\ll T_{K}$, the SR-LDOS for the FM island with an adatom following the procedure
found in Ref. [\onlinecite{GFM}], which was applied in the Kondo model with unpolarized bulk electrons. Such method allows to express
the SR-LDOS in terms of the phase shifts due to the adatom scattering and the Fano effect. This latter is originated in the interference
between the electron paths formed by the host conduction band and the Anderson impurity. We emphasize that the STM probe is not present
in this section. Initially we derive a formalism for finite $T$ and later on we take the limit $T \to 0$.
It is well known that
\begin{equation}
\rho_{LDOS}^{\sigma}\left(\omega,R\right)=-\frac{1}{\pi}\Im\left\{ \mathcal{G}_{\sigma}\left(\omega,R\right)\right\}, \label{eq:FM_LDOS}
\end{equation}
provides the SR-LDOS formula. The GF $\mathcal{G}_{\sigma}\left(\omega,R\right)$ is obtained from the Fourier transform of
\begin{equation}
\mathcal{G}_{\sigma}\left(\tau,R\right)=-\frac{i}{\hbar}\theta\left(\tau\right)Tr\left\{ \varrho_{FM}\left[\Psi_{R\sigma}\left(\tau\right),\Psi_{R\sigma}^{\dagger}\left(0\right)\right]_{+}\right\}\label{eq:G_PSI_PSI_}
\end{equation}
in time coordinate, where $\varrho_{FM}$ and $\left[,\right]_{+}$ are the density matrix of the FM island Hamiltonian and the anticommutator between the operators given by Eq. (\ref{eq:PSI_R}) at different times, respectively.

In order to be explicit, we have to apply the equation of motion procedure (EOM) on Eq. (\ref{eq:G_PSI_PSI_})
to demonstrate that $\mathcal{G}_{\sigma}\left(\omega,R\right)$ is coupled to other GFs as follows
\begin{equation}
\mathcal{G}_{\sigma}\left(\omega,R\right)=g_{\sigma}\left(\omega,0\right)+\tilde{g}_{\sigma}\left(\omega,R\right)\mathcal{T}_{\sigma}\left(\omega\right)\tilde{g}_{\sigma}\left(\omega,-R\right) . \label{eq:gf1}\end{equation}
The first term
\begin{equation}
g_{\sigma}\left(\omega,R\right)=\frac{1}{N_{FM\sigma}}\sum_{\vec{k}}\frac{e^{i\vec{k}.\vec{R}}}{\omega-\varepsilon_{k\sigma}+i\eta} , \label{eq:non_int_}\end{equation}
describes the bare GF for an uncorrelated electron state at the site $\vec{R}$, laterally displaced from the adatom and
\begin{equation}
\tilde{g}_{\sigma}\left(\omega,R\right)=\frac{1}{N_{FM\sigma}}\sum_{\vec{k}}\frac{\Gamma^{2}}{\Gamma^{2}+\varepsilon_{k\sigma}^{2}}\frac{e^{i\vec{k}.\vec{R}}}{\omega-\varepsilon_{k\sigma}+i\eta}\label{eq:non_int_2}\end{equation}
is the correspondent one dressed by the nonlocal hybridization [Eq. (\ref{eq:L})]. As a scattering center, the adatom defines a scattering amplitude
\begin{equation}
\mathcal{T}_{\sigma}\left(\omega\right)=\frac{\Delta}{\pi\rho_{0}}\mathcal{G}_{\sigma}^{dd}\left(\omega\right)\label{eq:scat_ampl}\end{equation}
proportional to the GF $\mathcal{G}{}_{\sigma}^{dd}\left(\omega\right)$.

The emergence of Fano interference and Friedel oscillations in the system SR-LDOS is a result of the interplay between Eqs. (\ref{eq:non_int_}) and (\ref{eq:non_int_2}). These effects can be elucidated by regrouping the terms in Eq. (\ref{eq:gf1}) in such way to achieve the form
$$
\rho_{LDOS}^{\sigma}\left(\omega,R\right)=\rho_{FM\sigma}+\pi\rho_{0}^{2}\left\{ \left(A_{\sigma}^{2}\left(R\right)-q_{FM\sigma}^{2}\right)\right.\nonumber \\
$$
\begin{equation}
\times \Im\left\{ \mathcal{T}_{\sigma}\left(\omega\right)\right\} +\left.2A_{\sigma}\left(R\right)q_{FM\sigma}\Re\left\{ \mathcal{T}_{\sigma}\left(\omega\right)\right\} \right\},\nonumber \\ \label{eq:free_LDOS}\end{equation}
where $\Re$ means real part,
\begin{equation}
q_{FM\sigma}=\frac{1}{\pi\rho_{0}}\Re\left\{\tilde{g}_{\sigma}\left(\omega,R\right)\right\}=\frac{\rho_{FM\sigma}}{\rho_{0}}J_{0}\left(k_{F\sigma}R\right)\frac{\Gamma}{\Gamma^{2}+\omega^{2}}\omega\label{eq:FM_Fano}\end{equation}
represents the Fano parameter due to the adatom-island hybridization in the wide-band limit and characterized by spin-polarized Friedel oscillations in the zeroth-order Bessel function $J_{0}\left(k_{F\sigma}R\right)$.
The spin-dependent Fermi wave numbers $k_{F\uparrow}$ and $k_{F\uparrow}$ are related to each other via
\begin{equation}
k_{F\uparrow}=\sqrt{\frac{1-P}{1+P}}k_{F\downarrow},\label{eq:ks}\end{equation}
deduced from Eqs. (\ref{eq:energy}) and (\ref{eq:rho_spin}). The SR-LDOS also depends on the function
\begin{equation}
A_{\sigma}\left(R\right)=\frac{1}{\pi\rho_{0}}\Im\left\{ \tilde{g}_{\sigma}\left(\omega,R\right)\right\} =\frac{\rho_{FM\sigma}}{\rho_{0}}J_{0}\left(k_{F\sigma}R\right)\frac{\Gamma^{2}}{\Gamma^{2}+\omega^{2}}\label{eq:Friedel}\end{equation}
that encloses spin-polarized Friedel oscillations for charge. Details of the method employed to derive Eq. (\ref{eq:FM_Fano}) appear in Appendix \ref{ApB}. We mention that to compare the SR-LDOS simulations of Sec. \ref{sec5} to the experimental data measured in a Fe island with a Co adatom \cite{Kawahara2010}, it is necessary  to set the Fano parameter in Eq. (\ref{eq:FM_Fano}), as a spinless ratio and equal to zero.

At low temperature limit, the SR-LDOS is well described by Eq. (\ref{eq:free_LDOS}) at the ground state. In such region, it is possible to classify the SR-LDOS
in terms of the spin-dependent phase shifts $\delta_{\sigma}\left(\omega\right)$ for the conduction electrons due to the adatom scattering center.
According to the results on the SIAM \cite{Langreth}, this characterization is obtained using the relation
\begin{eqnarray}
\exp\left[2i\delta_{\sigma}\left(\omega\right)\right] & =1-2\pi\rho_{0}\frac{1}{N_{FM\sigma}}\sum_{\vec{k}}\left(\frac{\Gamma^{2}}{\Gamma^{2}+\varepsilon_{k\sigma}^{2}}\right)^{2}\mathcal{T}_{\sigma}\left(\omega\right)i\nonumber \\
 & \times\delta\left(\omega-\varepsilon_{k\sigma}\right)\label{eq:phase_1}\end{eqnarray}
that correlates the phase shift $\delta_{\sigma}\left(\omega\right)$ to the real and imaginary parts of the scattering amplitude $\mathcal{T}_{\sigma}\left(\omega\right)$, given by Eq. (\ref{eq:scat_ampl}), thus resulting in the formula
\begin{equation}
\tan\delta_{\sigma}\left(\omega\right)=\frac{\Im\left\{ \mathcal{T}_{\sigma}\left(\omega\right)\right\} }{\Re\left\{ \mathcal{T}_{\sigma}\left(\omega\right)\right\} }.\label{eq:identity}\end{equation}
For the Fano interference we define an analogous relation, \cite{QD11,QD12,QD13} introducing the phase shift $\delta_{{q}_{FM}}$ as
\begin{equation}
\tan\delta_{q_{FM}}=-\frac{\Re\left\{ \tilde{g}_{\sigma}\left(\omega,R\right)\right\} }{\Im\left\{ \tilde{g}_{\sigma}\left(\omega,R\right)\right\} }=-\pi\rho_{0}\frac{q_{FM\sigma}}{\Im\left\{ \tilde{g}_{\sigma}\left(\omega,R\right)\right\} } , \label{eq:Fano_cb}\end{equation}
in order to show that
\begin{equation}
\rho_{LDOS}^{\sigma}\left(\omega,R\right)=\rho_{FM\sigma}\left[1-J_{0}^{2}\left(k_{F\sigma}R\right)\mathcal{F}_{\sigma}\left(\omega\right)\right] , \label{eq:free_LDOS_II}\end{equation}
becomes the expression for the system SR-LDOS and characterized by
\begin{equation}
\mathcal{F}_{\sigma}\left(\omega\right)=1-\frac{\cos^{2}\left(\delta_{\sigma}\left(\omega\right)-\delta_{q_{FM}}\right)}{\cos^{2}\delta_{q_{FM}}} , \label{eq:phases}\end{equation}
exclusively expressed in terms of the adatom scattering and Fano phase shifts. The last formula is the main result of this section, it represents the SR-LDOS of a metallic surface considered as a spin-polarized 2DEG coupled via a nonlocal hybridization to an adatom, in the framework of the SIAM given by Eq. (\ref{eq:2}). In Sec. \ref{sec3B}, we shall see that the STM spin-resolved conductance formula, assumes identical structures to Eqs. (\ref{eq:free_LDOS_II}) and (\ref{eq:phases}), but with a redefined Fano parameter due to the probe presence.

\subsection{Differential conductance and the SR-LDOS probed by the STM}
\label{sec3B}

In the low temperature regime and for a STM probe not very close to the metallic sample, the spin-resolved and differential conductance of the
STM device is proportional to an effective SR-LDOS that obeys the forms established by Eqs. (\ref{eq:free_LDOS_II}) and (\ref{eq:phases}).

To derive this effective SR-LDOS, we have to implement the linear response theory treating the tunneling Hamiltonian in Eq. (\ref{eq:4})
as a perturbation, just to ensure the weak tunneling regime as verified in experimental conditions \cite{Kawahara2010}.
Taking these assumptions into account we show that the spin-current follows the expression
\begin{equation}
I_{\sigma}=\frac{2e}{h}\pi\Gamma_{w}\int d\omega\left[n_{F}\left(\omega-e\phi\right)-n_{F}\left(\omega\right)\right]\tilde{\rho}_{LDOS}^{\sigma}\left(\omega\right) , \label{eq:current_III}\end{equation}
with $\Gamma_{w}=2\pi\left|w\right|^{2}\rho_{0}$ as the parameter that hybridizes the system conduction bands, $e$ the electron charge, $\phi$ the bias-voltage and
\begin{equation}
\tilde{\rho}_{LDOS}^{\sigma}\left(\omega,R\right)=-\frac{1}{\pi}\Im\left\{ \tilde{\mathcal{G}}_{\sigma}\left(\omega,R\right)\right\} \label{eq:LDOS}\end{equation}
as the effective SR-LDOS probed by the STM device. Such density is calculated using the GF $\tilde{\mathcal{G}}_{\sigma}\left(\omega,R\right)$ obtained from the time Fourier transform of
\begin{equation}
\tilde{\mathcal{G}}_{\sigma}\left(\tau,R\right)=-\frac{i}{\hbar}\theta\left(\tau\right)Tr\left\{ \varrho_{FM}\left[\tilde{\Psi}_{R\sigma}\left(\tau\right),\tilde{\Psi}_{R\sigma}^{\dagger}\left(0\right)\right]_{+}\right\} , \label{eq:G_PSI_PSI_2}\end{equation}
written in terms of the operator
\begin{equation}
\tilde{\Psi}_{R\sigma}=\Psi_{R\sigma}+\left(\pi\Delta\rho_{0}\right)^{1/2}q_{R}d_{\sigma} , \label{eq:O}\end{equation}
that describes the couplings between the probe and the island with the adatom, which depends on the new Fano parameter defined by
\begin{equation}
q_{R}=\left(\pi\Delta\rho_{0}\right)^{-1/2}\left(t_{dR}/w\right)=q_{o}\exp\left(-k_{F}R\right),\label{eq:Fano_II}\end{equation}
due to the interference between these additional conduction channels. To obtain the differential conductance $G_{\sigma}=\frac{\partial}{\partial \phi}I_{\sigma}$ for a given spin we consider Eq. (\ref{eq:current_III}) and show that
\begin{equation}
G_{\sigma}=\frac{2e^{2}}{h}\pi\Gamma_{w}\int d\omega\left\{ -\frac{\partial}{\partial\omega}n_{F}\left(\omega-e\phi\right)\right\} \tilde{\rho}_{LDOS}^{\sigma}\left(\omega,R\right)\label{eq:G_sigma_II}\end{equation}
is the spin-resolved conductance for the STM device.

Note that the spin effects on the system conductance lie on the free density of states of the island in Eq. (\ref{eq:rho_spin})
and on the scattering amplitude $\mathcal{T}_{\sigma}\left(\omega\right)$ in Eq. (\ref{eq:scat_ampl}) due to the adatom.
Thus, we need to express Eq. (\ref{eq:LDOS}) in terms of the adatom GF $\mathcal{G}_{\sigma}^{dd}\left(\omega\right)$ by employing the EOM procedure. This method leads to
\begin{eqnarray}
\tilde{\mathcal{G}}_{\sigma}\left(\omega,R\right) & = & \mathcal{G}_{\sigma}\left(\omega,R\right)+\pi\Delta\rho_{0}q_{R}^{2}\mathcal{G}_{\sigma}^{dd}\left(\omega\right)+\left(\pi\Delta\rho_{0}\right)^{1/2}q_{R}\nonumber \\
 & \times & \mathcal{G}_{\sigma}^{\Psi d}\left(\omega,R\right)+\left(\pi\Delta\rho_{0}\right)^{1/2}q_{R}\mathcal{G}_{\sigma}^{d\Psi}\left(\omega,R\right) , \label{eq:G_OO_I}\end{eqnarray}
and displays that such GF is linked to the Eq. (\ref{eq:gf1}) for the site $\vec{R}$ of the island and simultaneously to
\begin{equation}
\label{eq:G_PSI_d}
\mathcal{G}_{\sigma}^{\Psi d}\left(\omega,R\right)  =  \left(\pi\Delta\rho_{0}\right)^{1/2}\left[{q}_{FM\sigma}-iA_{\sigma}\left(R\right)\right]\mathcal{G}_{\sigma}^{dd}\left(\omega\right) , \end{equation}
and $\mathcal{G}_{\sigma}^{d\Psi}\left(\omega,R\right)$, where the former expression is obtained from the time Fourier transformation of
\begin{equation}
\mathcal{G}_{\sigma}^{\Psi d}\left(\tau\right)=-\frac{i}{\hbar}\theta\left(\tau\right)Tr\left\{ \varrho_{FM}\left[\Psi_{R\sigma}\left(\tau\right),d_{\sigma}^{\dagger}\left(0\right)\right]_{+}\right\},\label{eq:G_PSI_d_}\end{equation}
which is also equal to  $\mathcal{G}_{\sigma}^{\Psi d}\left(\omega,R\right)$. Thus replacing Eqs. (\ref{eq:scat_ampl}), (\ref{eq:G_OO_I}) and
\begin{eqnarray}
-\frac{1}{\pi}\Im\left\{ \mathcal{G}_{\sigma}^{d\Psi}\left(\omega,R\right)\right\}  & = & \left(\frac{\Delta\rho_{0}}{\pi}\right)^{1/2}\left\{A_{\sigma}\left(R\right)\Re\left\{ \mathcal{G}_{\sigma}^{dd}\left(\omega\right)\right\} \right.\nonumber \\
 & - & \left.{q}_{FM\sigma}\Im\left\{ \mathcal{G}_{\sigma}^{dd}\left(\omega\right)\right\} \right\} , \label{eq:Im_G_D_PSI}  \end{eqnarray}
into Eq. (\ref{eq:LDOS}), we find that
\begin{eqnarray}
\tilde{\rho}_{LDOS}^{\sigma}\left(\omega,R\right) & = & \rho_{FM\sigma}+\pi\rho_{0}^{2}\left\{ \left(A_{\sigma}^{2}\left(R\right)-q_{R\sigma}^{2}\right)\right.\nonumber \\
 & \times & \Im\left\{ \mathcal{T}_{\sigma}\left(\omega\right)\right\} +\left.2A_{\sigma}\left(R\right)q_{R\sigma}\Re\left\{ \mathcal{T}_{\sigma}\left(\omega\right)\right\} \right\} \nonumber \\ \label{eq:LDOS_II}\end{eqnarray}
obeys Eq. (\ref{eq:free_LDOS}) for the SR-LDOS in the case of an absent STM probe, but with an effective Fano parameter
\begin{equation}
q_{R\sigma}={q}_{FM\sigma}+{q}_{R} , \label{eq:Fano_ratio}\end{equation}
that takes into account an intrinsic Fano interference ($q_{FM\sigma}$) due to the adatom-island coupling represented by the first term and an extrinsic one ($q_R$) as a result of the STM probe hybridized with both adatom and the metallic surface,
enclosed by the second part. We pointed out that for $T\ll T_{K}$, the spin-resolved conductance in Eq. (\ref{eq:G_sigma_II}) becomes directly proportional to the effective SR-LDOS evaluated at the energy $e\phi$. To show that, we use the Dirac delta distribution expressed by the minus derivative of the Fermi function in Eq. (\ref{eq:G_sigma_II}), which eliminates the integration over energy and gives
\begin{equation}
G_{\sigma}=\frac{2e^{2}}{h}\pi\Gamma_{w}\tilde{\rho}_{LDOS}^{\sigma}\left(e\phi\right).\label{eq:G_sigma_III}\end{equation} This result means that the effective SR-LDOS is a fairly representative function for the spin-resolved conductance of the system, which favors us to apply the zero temperature formalism of phase shifts discussed in Sec. \ref{sec3A} by introducing
\begin{equation}
\tan\delta_{q_{R\sigma}}=-\frac{q_{R\sigma}}{A_{\sigma}\left(R\right)} , \label{eq:Fano_phase_II}\end{equation}
as the total Fano phase shift. Combining it with Eq. (\ref{eq:identity}) for the scattering amplitude $\mathcal{T}_{\sigma}\left(\omega\right)$, it is possible to derive
\begin{equation}
\tilde{\rho}_{LDOS}^{\sigma}\left(\omega,R\right)=\rho_{FM\sigma}\left[1-J_{0}^{2}\left(k_{F\sigma}R\right)\tilde{\mathcal{F}}_{\sigma}\left(\omega,R\right)\right]\label{eq:Free_LDOS_III}\end{equation}
and
\begin{equation}
\tilde{\mathcal{F}}_{\sigma}\left(\omega,R\right)=1-\frac{\cos^{2}\left(\delta_{\sigma}\left(\omega\right)-\delta_{{q}_{R\sigma}}\right)}{\cos^{2}\delta_{{q}_{R\sigma}}} , \label{eq:phases_}\end{equation}
as expressions that represent the effective SR-LDOS probed by the STM. The latter contains two sources for Fano effect: the first concerns on the Fano interference between the traveling electrons through the host conduction band that can ``visit" the adatom site and go back to it, and those that do not perform such ``visit". Additionally, the second process is composed by the couplings of the probe with the island and the adatom. It also has the scattering of the traveling electrons due to the side-coupled adatom, which in certain conditions leads to the Kondo effect. We also remark that the phase shift $\delta_{\sigma}\left(\omega\right)$ given by Eq. (\ref{eq:identity}), is obtained in this work employing the atomic approach.

From Eq. (\ref{eq:Free_LDOS_III}), we are able to determine the following occupation number
\begin{equation}
n_{LDOS}^{\sigma}=\int_{-\infty }^{+\infty }d\omega \tilde{\rho}_{LDOS}^{\sigma}(\omega,R=0)n_{F}(\omega).\label{nSR_LDOS}
\end{equation}
As we shall see, this formula will guide us to better understand the results of Sec. \ref{sec4}.

\section{RESULTS}
\label{sec4}

Here we present the results obtained via the formulation developed in the previous section.
We employ a set of parameters that defines the Kondo regime: $E_{d}=-10.0\Delta$ and $V=12.0\Delta$,
where $\Delta=0.01D_{0}$ $(D_{0}=1.0)$. We will basically compare two regimes of $q_0$ value.
The large $q_0$ limit we call \emph{peak limit} due to the formation of a Kondo peak in the LDOS.
The other regime, corresponding to small and intermediate values of $q_0$, we call
\emph{dip limit} since the LDOS presents a dip around the Fermi level.

It is necessary to mention that the SR-LDOS formula derived in the previous section is valid for zero temperature.
On the other hand the numerical procedure of the atomic approach is well established for $T\ll T_{K}$ but still finite $T$.
So we consider a very low temperature $T=0.001\Delta$ to allow the combination of both procedures.
To investigate the spatial behaviors of the Fano factor and the LDOS,
we define a dimensionless parameter $k_{F\downarrow}R=k_{F}R$, to represent the STM probe-adatom lateral distance in Fig. \ref{fig:Pic1}.

\subsection{Intrinsic Fano parameter}
\label{sec4NLC}

\begin{figure}[h]
% \vskip0.5cm
\centerline{\resizebox{3.5in}{!}{
\includegraphics[clip,width=0.40\textwidth,angle=-90.]{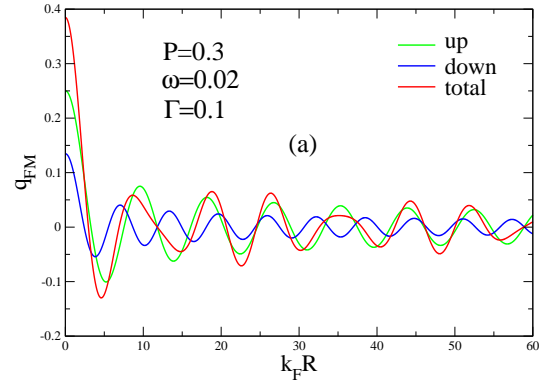}}}
\centerline{\resizebox{3.5in}{!}{
\includegraphics[clip,width=0.40\textwidth,angle=-90.]{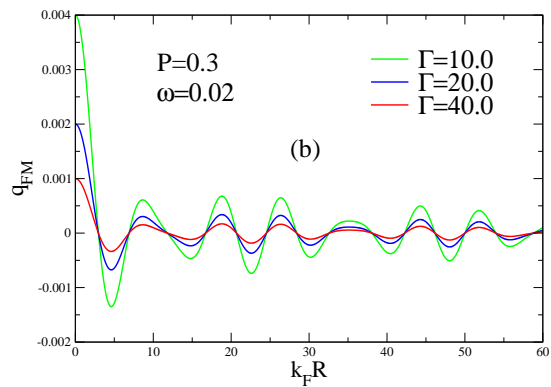}}}
\caption{(a) Spin-polarized Fano factors $q_{FM\sigma}$ for $\Gamma=0.1$ and
(b) the sum $q_{FM}=q_{FM\uparrow}+q_{FM\downarrow}$ for $\Gamma=10.0$, $\Gamma=20.0$ and $\Gamma=40.0$
in units of $\omega$ as a function of the dimensionless parameter $k_{F}R$.}\label{fano}
\end{figure}

Here we look with some more detail the spatial dependence of the intrinsic Fano factor $q_{FM\sigma}$.
Fig. \ref{fano}(a) shows $q_{FM\sigma}$ against $k_F R$ for  $P=0.3$, $\omega=0.02D_{0}$
and $\Gamma=0.1\omega$. Interestingly, this Fano factor reveals spin-polarized Friedel oscillations.
These oscillations exhibit enhanced amplitudes for the spin-up channel and a phase
shifted pattern in relation to the spin-down component, due to the spin-dependent Fermi wave number [Eq. (\ref{eq:ks})].
This phase shift yields irregular oscillations in the total Fano parameter $q_{FM}=q_{FM\uparrow}+q_{FM\downarrow}$
[Figs. \ref{fano}(a) and \ref{fano}(b)].

In Fig. \ref{fano}(b) we can also observe by changing from $\Gamma=10\omega$ to $\Gamma=40\omega$, that the spin-polarized
Fano parameters approach to zero, exhibiting flattened oscillations as a function of $k_{F}R$. For large enough $\Gamma$, we have
$q_{FM\sigma} \approx 0$. Since the present atomic approach is valid only for constant $V_{d k \sigma}$ which is
obtained for large $\Gamma$ (local coupling), we will restrict our analysis to negligible
$q_{FM\sigma}$ values. In particular, for a perfect local coupling configuration, settled by the condition $q_{FM\sigma}=0$ in Eq. (\ref{eq:FM_Fano}),
we can conclude according to Eq. (\ref{eq:Fano_ratio}), that the Fano parameter $q_{o}$ induced by the STM probe is the only one
that rules the Fano interference in the system. In Secs. \ref{sec4LC} and \ref{sec4FKL}, we discuss the possible interference limits for $q_{o}$
in the local coupling regime, where the atomic approach is applicable.

\subsection{SR-LDOS in the peak limit ($q_o=100$)}
\label{sec4LC}

\begin{figure}[h]
% \vskip0.5cm
\centerline{\resizebox{3.2in}{!}{
\includegraphics[clip,width=0.40\textwidth,angle=-90.]{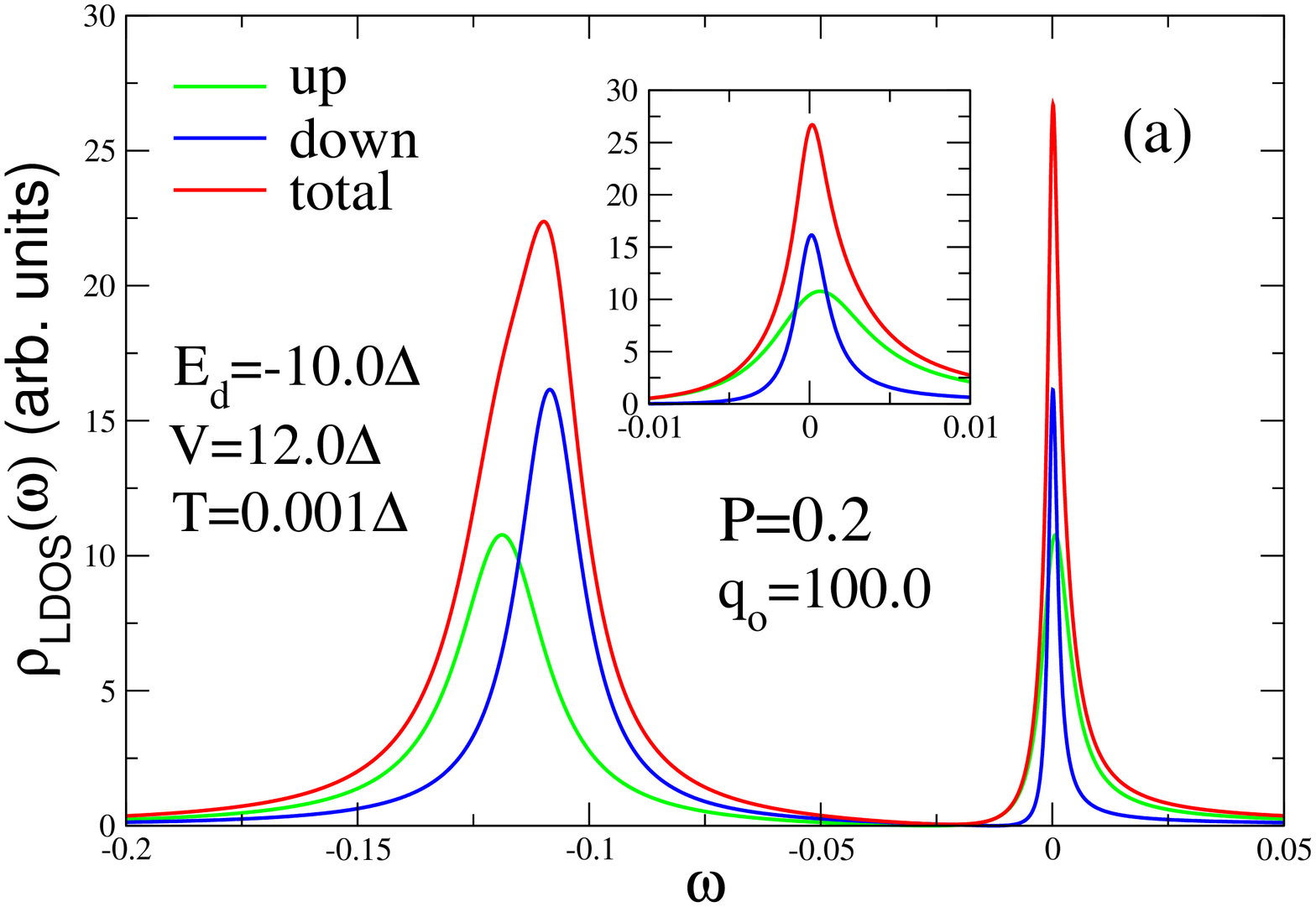}}}
\centerline{\resizebox{3.2in}{!}{
\includegraphics[clip,width=0.40\textwidth,angle=-90.]{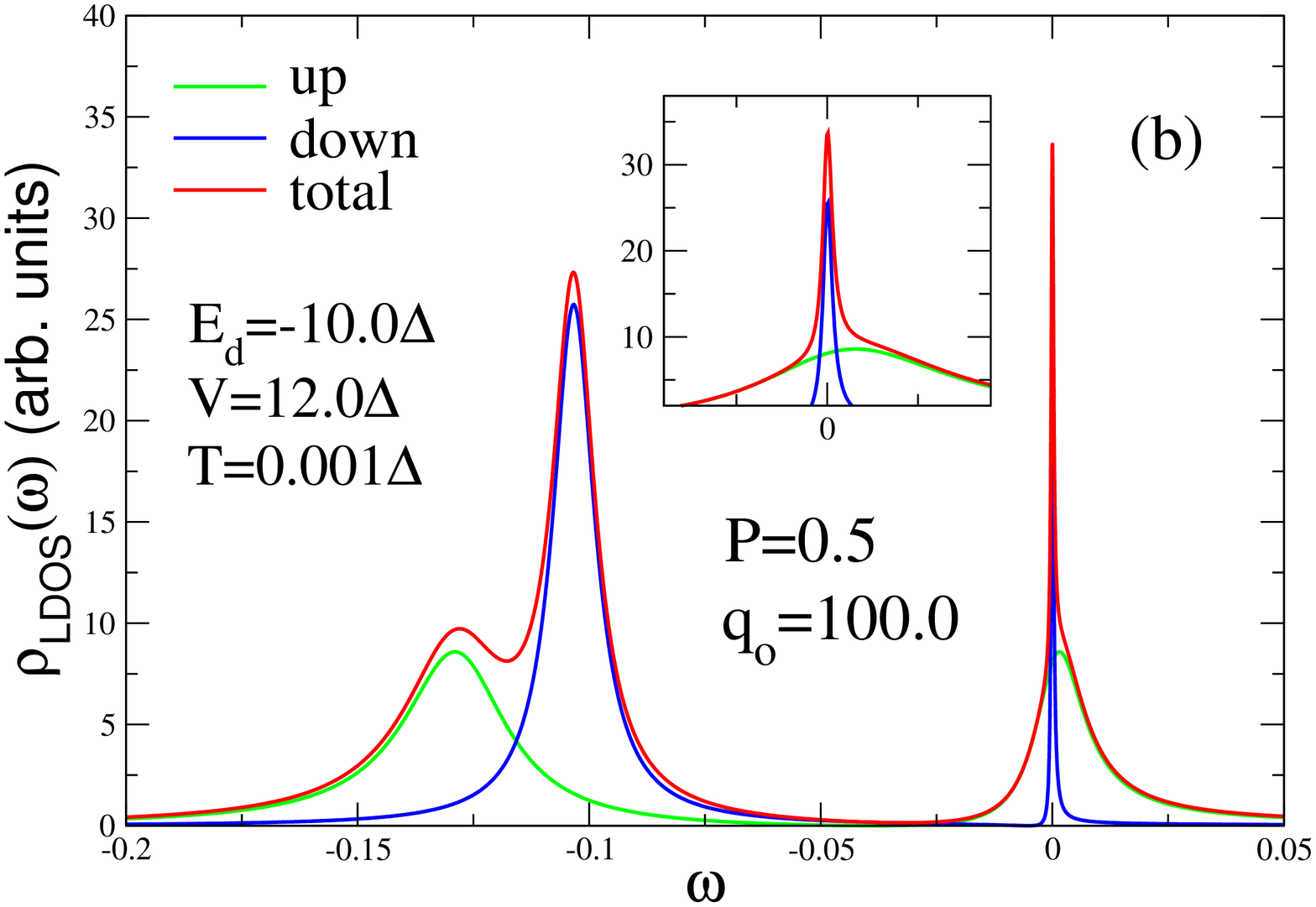}}}
\centerline{\resizebox{3.2in}{!}{
\includegraphics[clip,width=0.40\textwidth,angle=-90.]{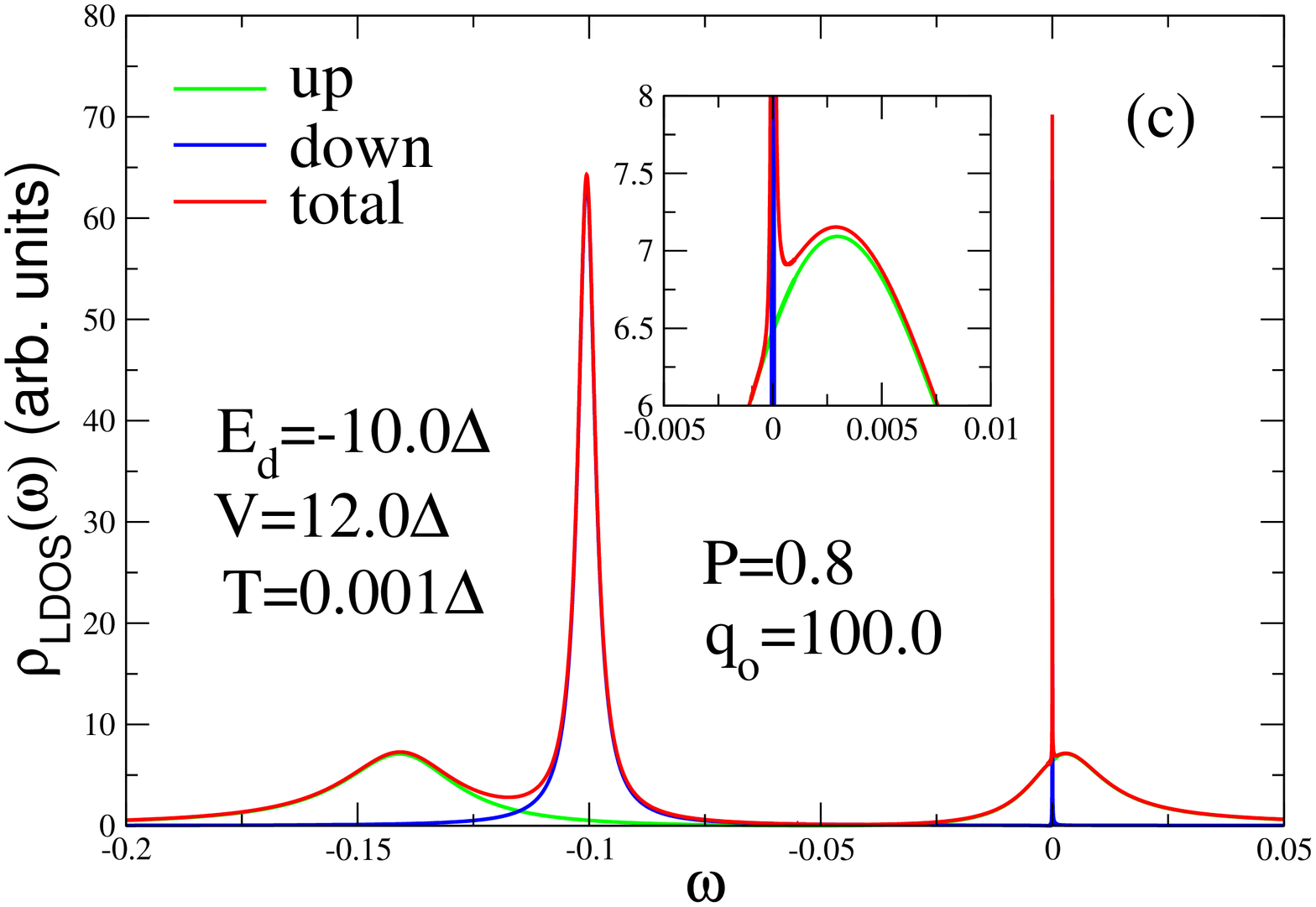}}}
\caption{Spin-resolved local density of states (SR-LDOS) at $R=0$, in arbitrary units (a.u.),
as a function of the energy $\omega$ for increasing $P$ value: (a) $P=0.2$, (b) $P=0.5$ and (c) $P=0.8$.
In the insets we show the SR-LDOS around the host Fermi level, where we can see the Kondo peak splitting
with the pinning of the spin down component. We also observe that as $P$ increases the spin down resonance
shrinks.}\label{fig678}
\end{figure}

Figs. \ref{fig678}(a)-(c) show the SR-LDOS and total LDOS for increasing polarization degree of the host.
In the unpolarized case $P=0$ (not shown) we have the well known twofold degenerate Kondo peak as in the SET.\cite{QD1,QD2}
However, for increasing $P$ this degeneracy is broken and the Kondo peak splits into two peaks. While the corresponding spin-up
peak presents a slight blue-shift for increasing $P$, the spin-down one remains pinned around the host Fermi level. Additionally, the
widths of these peaks show opposite behaviors. While the up-peak broadens the down-peak shrinks as $P$ enlarges.
In the insets we present with more detail the SR-LDOS around the Fermi level, where we can see more clearly the
spin-splitting of the Kondo resonance.

It is valid to mention that our spin-splitting resembles the work of Y. Qi \textit{et al.}\cite{Yunong}
for a spin current injected in a nonmagnetic conductor with a Kondo adatom. In this work
it is possible to note a tendency of a pinning for one particular spin component of the Kondo peak.
Kondo peak splitting was also observed in a QD system hybridized to ferromagnetic reservoirs\cite{FM2}
and in the presence of spin-flip\cite{FM1}.

In the experimental point of view the spin-splitting of the Kondo resonance has already
been observed in systems of QDs coupled to ferromagnetic reservoirs \cite{FM9} and in a carbon nanotube QD interacting
with a magnetic particle.\cite{Nygard2004} Observe that for $P=0.2$ and $P=0.5$ it is not
possible to resolve the spin-splitting of the Kondo peak in the total LDOS. Experimentally, non-resolved
Kondo peak splitting can also occur.\cite{FM11} Although the STM experiments are in general restricted to energies around the host Fermi level, here we show the SR-LDOS for a wider energy window
in order to see the polarization effects on the adatom level $E_{d}$, as displayed in Fig. \ref{fig678}.

\begin{figure}
\includegraphics[clip,width=0.40\textwidth,angle=-90.]{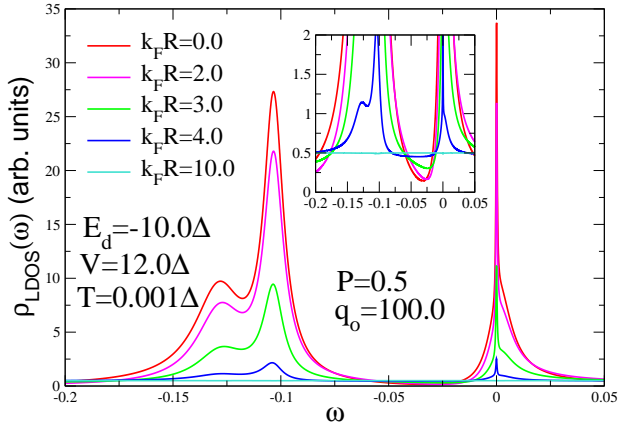}
\caption{Local density of states (LDOS), in arbitrary units (a.u.),  as a function of the energy $\omega$ for different
$k_{F}R$ values. In the inset, we present the LDOS in  detail.}
\label{Fig31}
\end{figure}

The analysis of the non-Zeeman splitting in the full LDOS for different STM probe positions is presented in Fig. \ref{Fig31} for $P=0.5$.
We see that the LDOS profile becomes flatter for large enough values of the dimensionless parameter $k_{F}R$, thus revealing a crossover
from the \emph{peak limit} at $R=0$ to the background value represented by the host free density of states [Eq. (\ref{eq:den})].

\begin{figure}
\includegraphics[clip,width=0.40\textwidth,angle=-90.0]{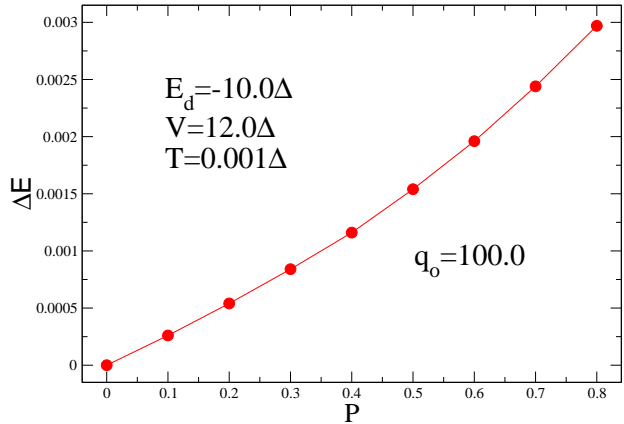}
\caption{Kondo peak splitting $\Delta E$ in units of $D_{0}$, as a function of the host polarization $P$.}
\label{Fig5}
\end{figure}

We end this section presenting Fig. \ref{Fig5}, where the dependence of the spin-splitting of the Kondo peak $\Delta E$
as a function of the host polarization $P$ can be observed. Note that  $\Delta E$ displays a nonlinear behavior as $P$ increases.
This nonlinearity was also found by Y. Qi \textit{et al.}\cite{Yunong}

\subsection{SR-LDOS in the dip limit $(q_0 \leq 1)$}
\label{sec4FKL}

\begin{figure}[h]
% \vskip0.5cm
\centerline{\resizebox{3.5in}{!}{
\includegraphics[clip,width=0.40\textwidth,angle=-90.]{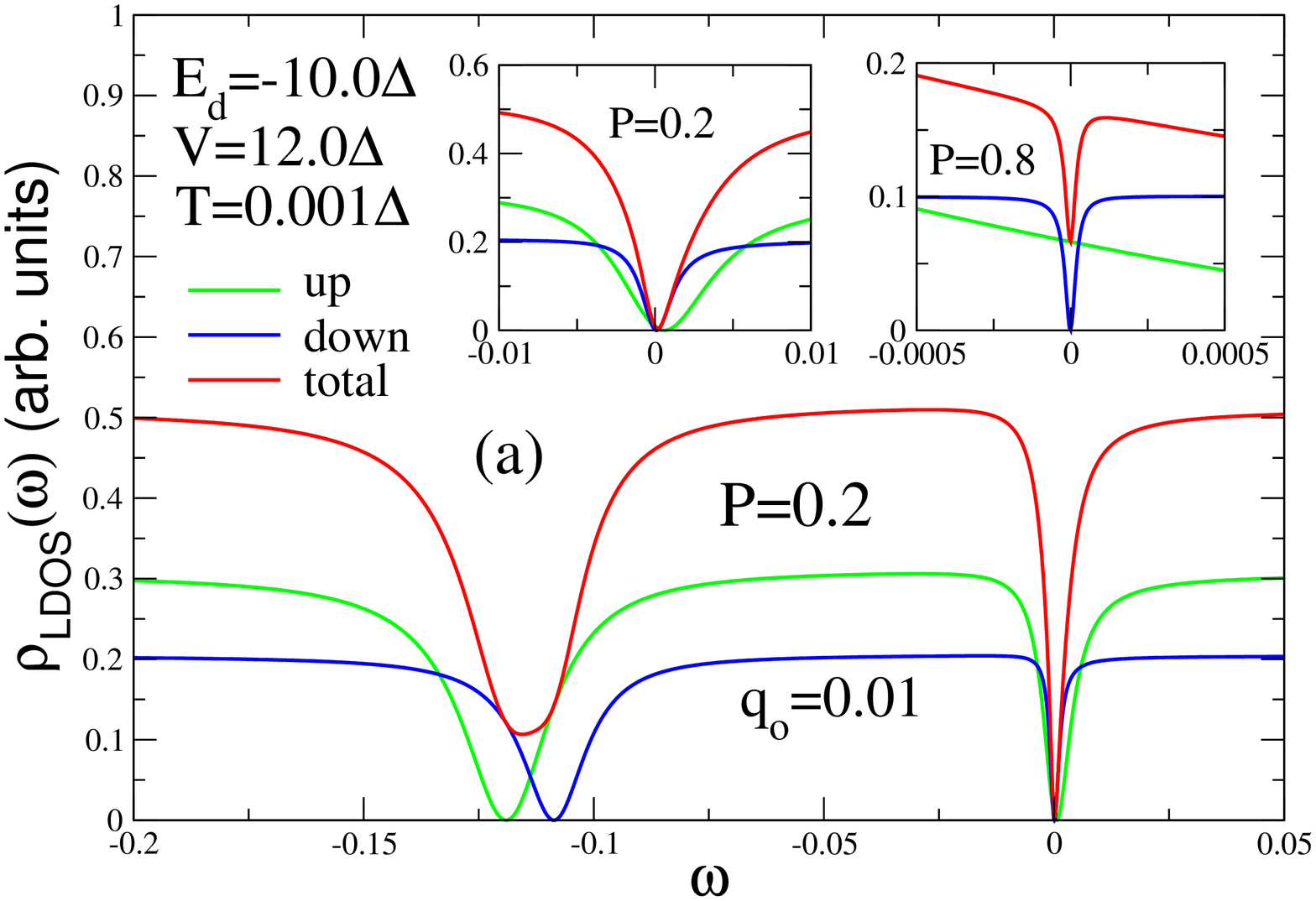}}}
\centerline{\resizebox{3.5in}{!}{
\includegraphics[clip,width=0.40\textwidth,angle=-90.]{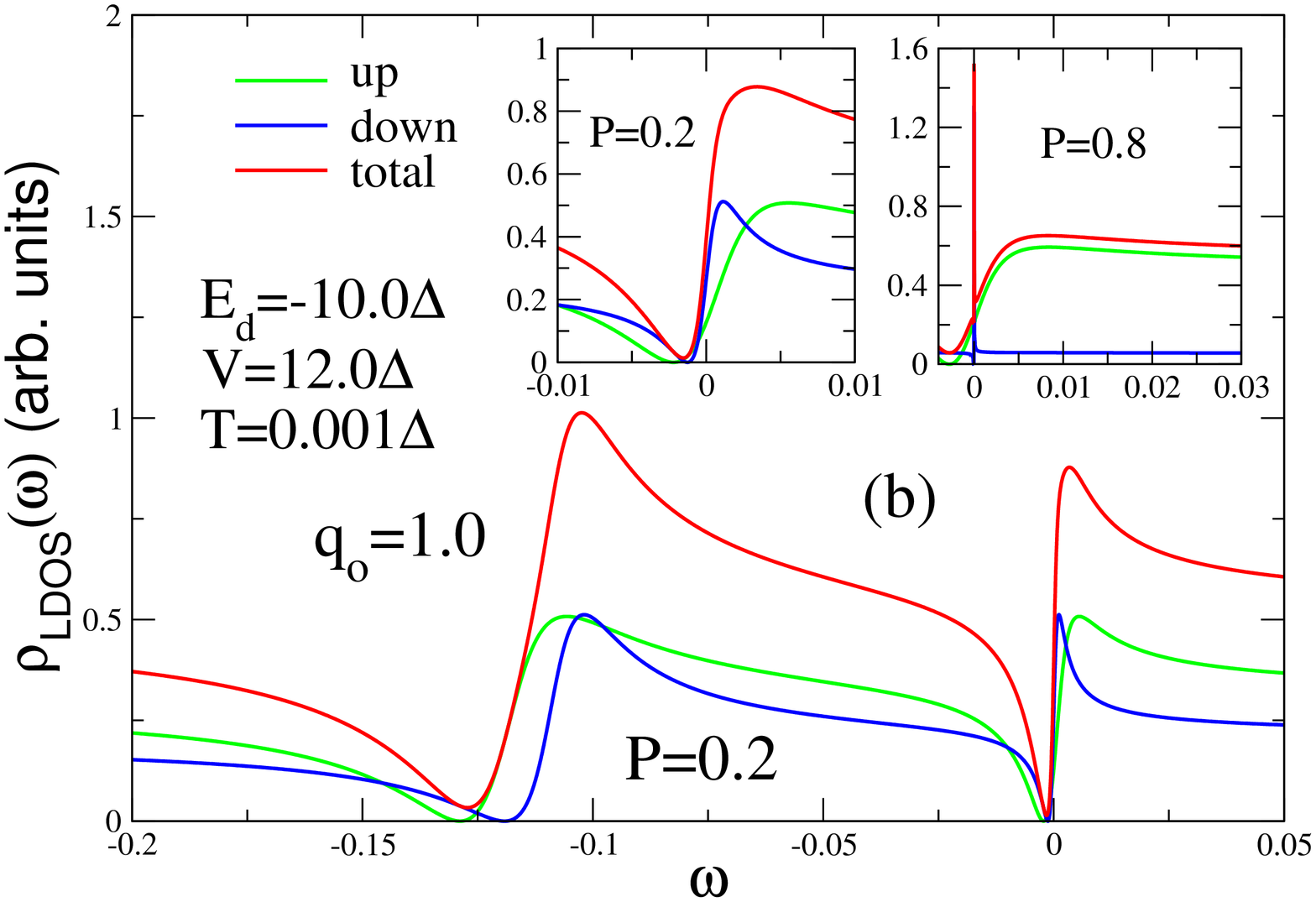}}}
\caption{Spin-resolved local density of states (SR-LDOS) at $R=0$, in arbitrary units (a.u.),  as a function of the energy $\omega$.
In the full curve we show the local density of states (LDOS). In the insets, we present the SR-LDOS and the LDOS in the vicinity of the
Fermi energy $\omega\simeq\varepsilon_{F}=0$. For $P=0.2$, there is a Fano-Kondo structure in the LDOS profile, but for $P=0.8$ an enhanced Kondo peak appears in the spin-down channel.}\label{dip1}
\end{figure}

In contrast to the $q_0=100$ case previously analyzed in the \emph{peak limit},
here we discuss the $q_0 \leq 1$ regime.
This corresponds to a direct STM-probe and host surface tunneling being dominant.
In this situation, the system reduces to an adatom side-coupled to the FM island as represented in Fig. \ref{fig:Pic1},
which is equivalent to the T-Shaped QD device \cite{QD7}.
For small Fano factors ($q_0=0.01$), the LDOS exhibits antiresonances (dips) instead of resonances,
as we can see in Fig. \ref{dip1}(a). This happens in such a way that the antiresonance in the up SR-LDOS channel,
disappears for high enough polarization as seen in the inset of Fig. \ref{dip1}(a). On the other hand, the dip of the down component
shrinks as $P$ increases. The width of this antiresonance lowers two orders of magnitude when we change the polarization from $P=0.2$ to $P=0.8$.
So we can conclude that the polarization induces a continuous second order insulator-metal transition in the system.
As the polarization grows from $P=0$ to $P=0.8$, the up SR-LDOS component becomes continuously flat, generating a finite SR-LDOS in the vicinity of
the Fermi energy $\omega\simeq\varepsilon_{F}=0$. At the same time, the down SR-LDOS component shrinks, closing the total LDOS gap, as indicated in
the inset of Fig. \ref{dip1}(a) for $P=0.8$.

The last cases in the \emph{dip limit} we investigate, corresponds to $q_{o}=1$, $P=0.2$ and $P=0.8$ as presented in Fig. \ref{dip1}(b). For the
low polarization $P=0.2$, the Fano interference is robust presenting the coexistence of a structure composed by dips and peaks (Fano line shape).
However, for the large polarization $P=0.8$, the Kondo peak in the adatom GF prevails in the spin-down channel and suppress the destructive interference.
This feature appears in the insets of Fig. \ref{dip1}(b).

The simulations presented resemble experimental results found in ferromagnetic contacts as in the M. R. Calvo \textit{et al.} work \cite{FM11}
(see their Fig. 1c) and adatoms systems discussed by N. N\'eel \textit{et al.}\cite{SPSTM5} (see their Fig. 1b). In the former experiment,
depending on the electrodes composition coupled to the contact, the energy profile of the conductance displays patterns without resolved spin-splitting.
This behavior  is related to the cases of small and intermediate values of the Fano parameter, which were obtained with Fe and Co contacts respectively.
For the second experiment, spin-polarized STM  probes made by Fe or W were employed on a Co adatom deposited on a Cu(111) surface. In particular,
in this case no resolved spin-splitting was verified, which corresponds to the limit of small Fano factor. These setups can be reproduced in our
simulations considering intermediate polarizations.

\subsection{Phase shifts and occupation numbers}
\label{sec4KL}

\begin{figure}[h]
% \vskip0.5cm
\centerline{\resizebox{3.5in}{!}{
\includegraphics[clip,width=0.40\textwidth,angle=-90.]{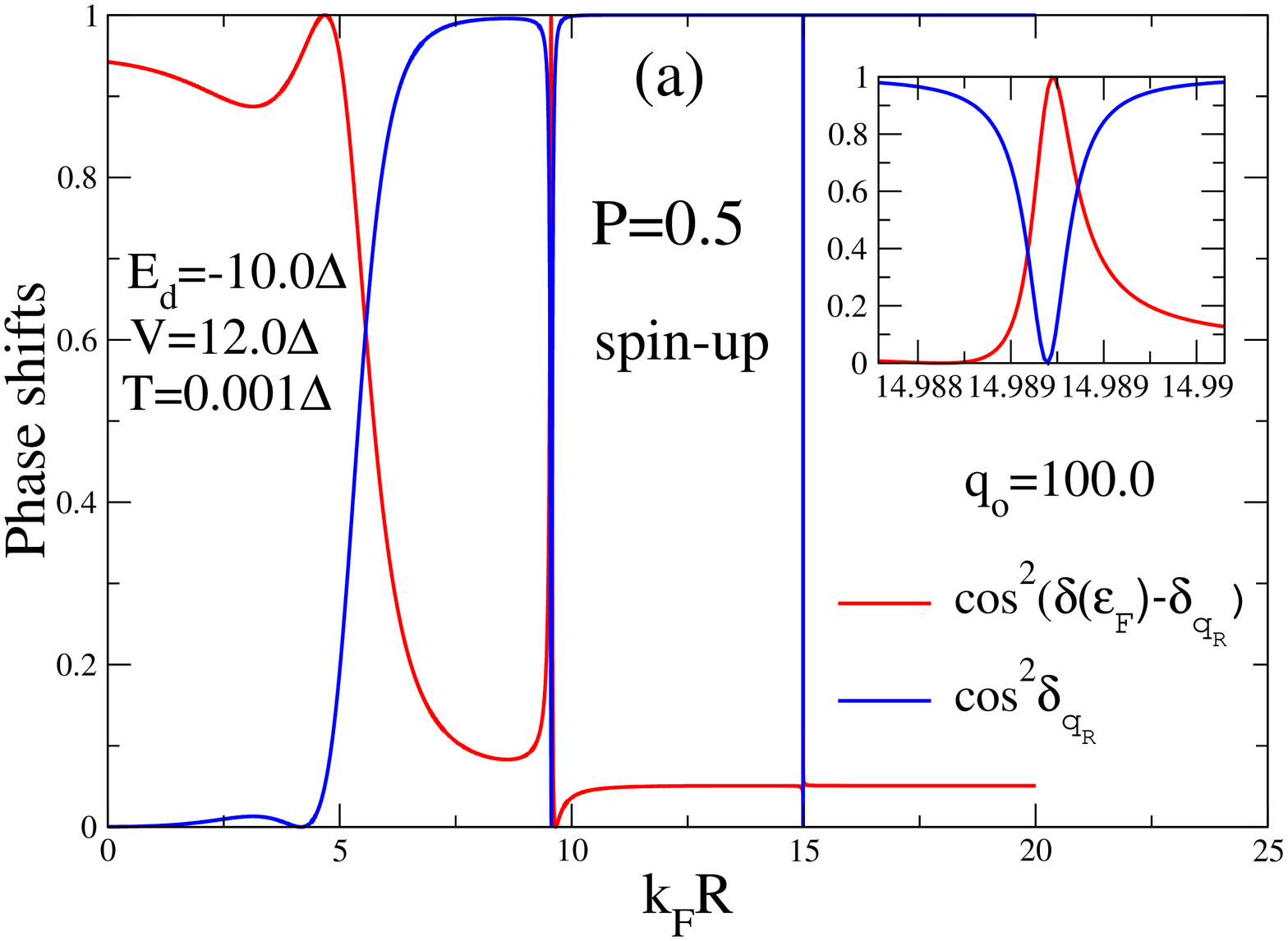}}}
\centerline{\resizebox{3.5in}{!}{
\includegraphics[clip,width=0.40\textwidth,angle=-90.]{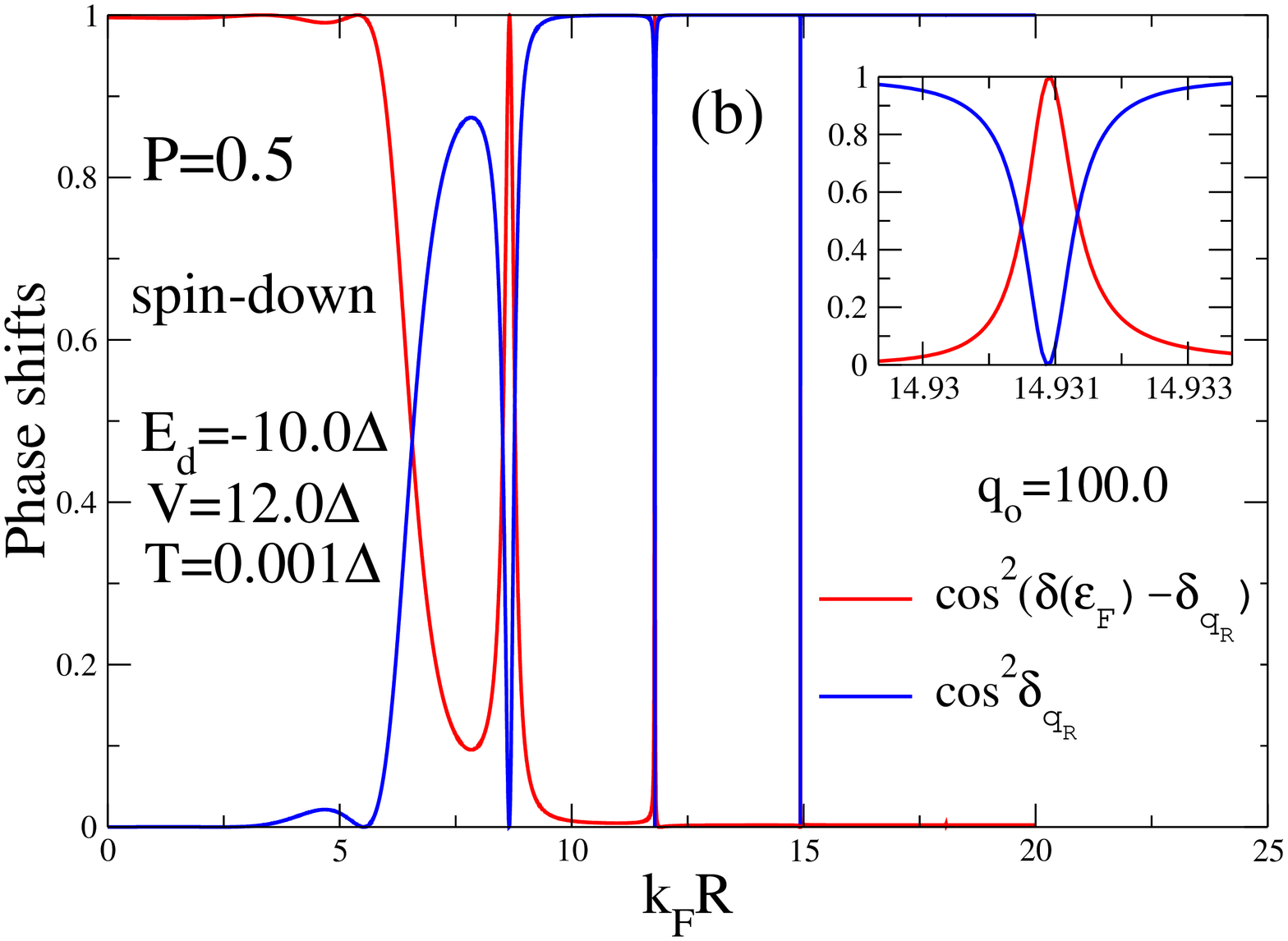}}}
\caption{Analysis of the system phase
shift structure evaluated at the host Fermi level. Plots of $\cos^{2}\left(\delta_{\sigma}\left(\varepsilon_{F}\right)-\delta_{q_{R\sigma}}\right)$ and $\cos^{2}\delta_{q_{R\sigma}}$,
as a function of the dimensionless parameter $k_{F}R$ for both spin components. In the inset, we show the resonance and antiresonance features at $k_{F}R \simeq 15.0$.}\label{phase}
\end{figure}

In Sec. \ref{sec3B} we show that the SR-LDOS can be expressed in terms of the phase shifts $\delta_{\sigma}\left(\omega\right)$ and $\delta_{{q}_{R\sigma}}$.
Here we explore the quantities $\cos^{2}\delta_{{q}_{R\sigma}}$ and $\cos^{2}\left(\delta_{\sigma}\left(\omega\right)-{\delta}_{{q}_{R\sigma}}\right)$
in order to gain further insight about the spin-splitting and Kondo peak pinning found in the SR-LDOS. In Fig. (\ref{phase}) we present these
two functions against $k_F R$ for both spin components. The numerical parameters are indicated in the plots and curves are evaluated
at the Fermi level. The main distinction between
the up and down curves can be observed in the range $0 < k_{F}R < 5.0$ where $\cos^{2}\left(\delta_{\sigma}\left(\varepsilon_{F}\right)-{\delta}_{{q}_{R\sigma}}\right) \approx 1$
for spin-down while it is suppressed for spin-up. In particular, for $k_F R \to 0$ we find $\cos^2 (\delta_{{q}_{R\sigma}}) \to 0 $
which means that $\delta_{{q}_{R\sigma}} \to \pi/2$. Consequently, $\cos^{2}\left(\delta_{\sigma}\left(\varepsilon_{F}\right)-{\delta}_{{q}_{R\sigma}}\right) \to \sin^2 (\delta_{\sigma}\left(\varepsilon_{F}\right))$.
On the other hand, according to Friedel sum rule [Eq. (\ref{fried})] we have $\delta_{\sigma}\left(\varepsilon_{F}\right) = \pi n_{d\sigma}$. This implies that $n_{d\downarrow} \to 0.5$ while
$n_{d\uparrow}$ stays below 0.5. The fact that $n_{d\downarrow}$ remains close to 0.5 results in the pinning of the spin-down component of the Kondo resonance.
For increasing $k_F R$ the curves obtained from Eq. (\ref{eq:Fano_phase_II}) display a series of peaks and dips due to the interplay between the exponential
decay in Eq. (\ref{eq:td}) and the Friedel oscillations.

\begin{figure}
\includegraphics[clip,width=0.40\textwidth,angle=-90.]{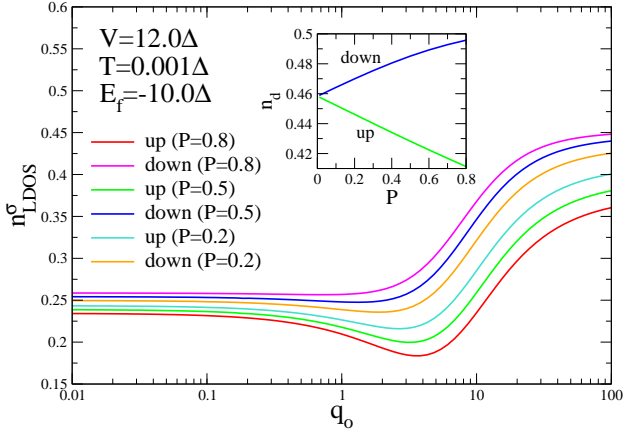}\caption{Occupation number $n^{\sigma}_{LDOS}$ of
Eq. (\ref{nSR_LDOS}) as a function of the Fano parameter $q_o$ and with different values for the spin-polarization $P$ of the host. In the inset, we present $n_{d\sigma}$ against $P$.}
\label{Fig70}
\end{figure}

To complement the analysis on the emergence of the Kondo peak pinning, we present Fig. \ref{Fig70}. In such figure, we plot the occupation number $n^{\sigma}_{LDOS}$ of Eq. (\ref{nSR_LDOS}) as a function of the Fano factor $q_{0}$ for differing spin-polarizations $P$. This gives also the same tendency of
$n_{d,\uparrow}$ and $n_{d,\downarrow}$, since both $n^{\sigma}_{LDOS}$ and $n_{d,\sigma}$ should preserve some proportionality.
In the large $q_0$ limit, $n^{\downarrow}_{LDOS}$ approaches to 0.45 while $n^{\uparrow}_{LDOS}$ moves to lower values as $P$ increases.
This corroborates the pinning of the spin-down Kondo resonance. Note that the Friedel sum rule given by Eq. (\ref{fried}),
ensures the pinning of the Kondo peak in the down-channel and the displacement of the up peak, via the inequality $n_{d,\uparrow} < n_{d,\downarrow} \approx 0.5$. In the inset of Fig. \ref{Fig70}, the plots of $n_{d,\uparrow}$ and $n_{d,\downarrow}$ against $P$ confirm this inequality. Such occupations indicate that there is a net
magnetic moment at the adatom partially screened by the conduction electrons, where the lack of spin-down conduction electrons are not able to blind
the spin-up component of the impurity, thus avoiding a formation of a defined Kondo peak in that channel. On the other hand, the excess of spin-up
conduction electrons yields to a Kondo peak in the down-component of the adatom DOS. We can conclude that, there is no abrupt break down of the Kondo effect,
but a crossover from the ordinary Kondo effect to a situation where the Kondo effect is gradually suppressed.

\section{Analysis of some experimental results}
\label{sec5}

In this section we successfully apply the present developed formulation to reproduce recent experimental findings on a single Kondo adatom coupled
to a magnetic cluster. To our best knowledge, the first experimental work that explores Kondo adatom on a ferromagnetic host was
recently done by S. L. Kawahara \textit{et al.} \cite{Kawahara2010}, which used an unpolarized STM probe
on top of a Fe island with a Co adatom. They observed that, depending on the adsorption site of the Co atom, a spin-splitting of
a Fano-Kondo dip is induced by the spin-polarization of the island, and can be explained by a double Fano antiresonance in
a single particle picture.

As pointed out in Ref. [\onlinecite{Kawahara2010}] there are two competing mechanisms that can result or
not in the Kondo effect. The first one is the antiferromagnetic exchange interaction between the Co adatom and the itinerant $sp$ island electrons.
The second one is the exchange interaction due to the direct $d-d$ ferromagnetic interaction of the Co adatom and the Fe atoms of the magnetic
substrate. Our formulation only can be applied to this system  if the antiferromagnetic interaction dominates over
the direct ferromagnetic correlation. This competing processes appear in Ref. [\onlinecite{Kawahara2010}] as an assumption, but further experimental
and theoretical investigations are necessary to better understand the dominant mechanism.

There are some related experimental results, supported by first principles calculations of M. R. Calvo \textit{et al.}\cite{FM11},
in related systems. They found the existence of a Kondo peak in ferromagnetic atomic contacts hybridized with electrodes
(both built by Fe, Ni and Co), differently from those found in the bulk limit. In  nanoscale, the electrons at
the surfaces of these junctions experience interactions where the antiferromagnetic coupling overcomes the ferromagnetic correlations.
In such setups, the nanocontacts play the role of the Co adatom used in the Kawahara \textit{et al} work. Thus, our model in its present
form does not support the opposite case characterized by strong ferromagnetic correlations, which are usually modeled by a Heisenberg type
interaction. For an enhanced antiferromagnetic coupling between the $sp$ electrons and the adatom, the picture of a spin-polarized electrons
gas as discussed by M. R. Calvo \textit{et al.}\cite{FM11} can be employed to describe a ferromagnetic metallic sample with a Kondo impurity. Additionally, we would like to remark that Kondo adatoms and some QD systems indeed have a spin $S>1/2$,
which can be detectable by a magnetic anisotropy signature .\cite{STM12,new3,new5}.
In these cases, a multi-orbital Anderson Hamiltonian could offer a more detailed modeling \cite{STM11,new1}
and improve the accuracy of the present work.

However, we changed the localized adatom level $E_{d}$ in all the relevant parameter range of
the SIAM in order to better reproduce the line shape of Kawahara \textit{et al.} work.
The optimized values are $E_{d}=-3.0\Delta$ and $q_{0}=0.01$, which characterizes the intermediate
valence regime of the system.

\begin{figure}[h]
% \vskip0.5cm
\centerline{\resizebox{3.5in}{!}{
\includegraphics[clip,width=0.40\textwidth,angle=-90.]{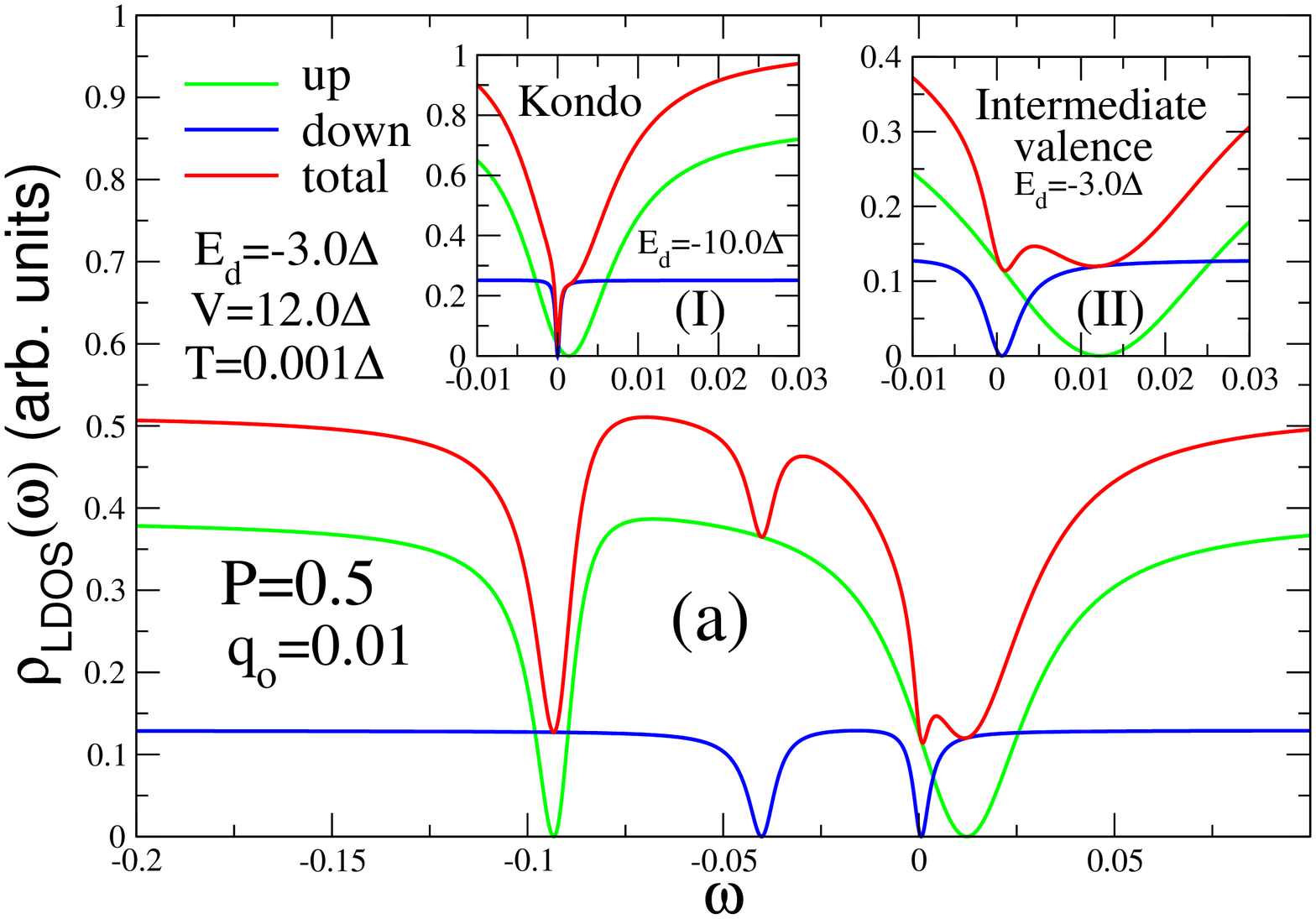}}}
\centerline{\resizebox{3.5in}{!}{
\includegraphics[clip,width=0.40\textwidth,angle=-90.]{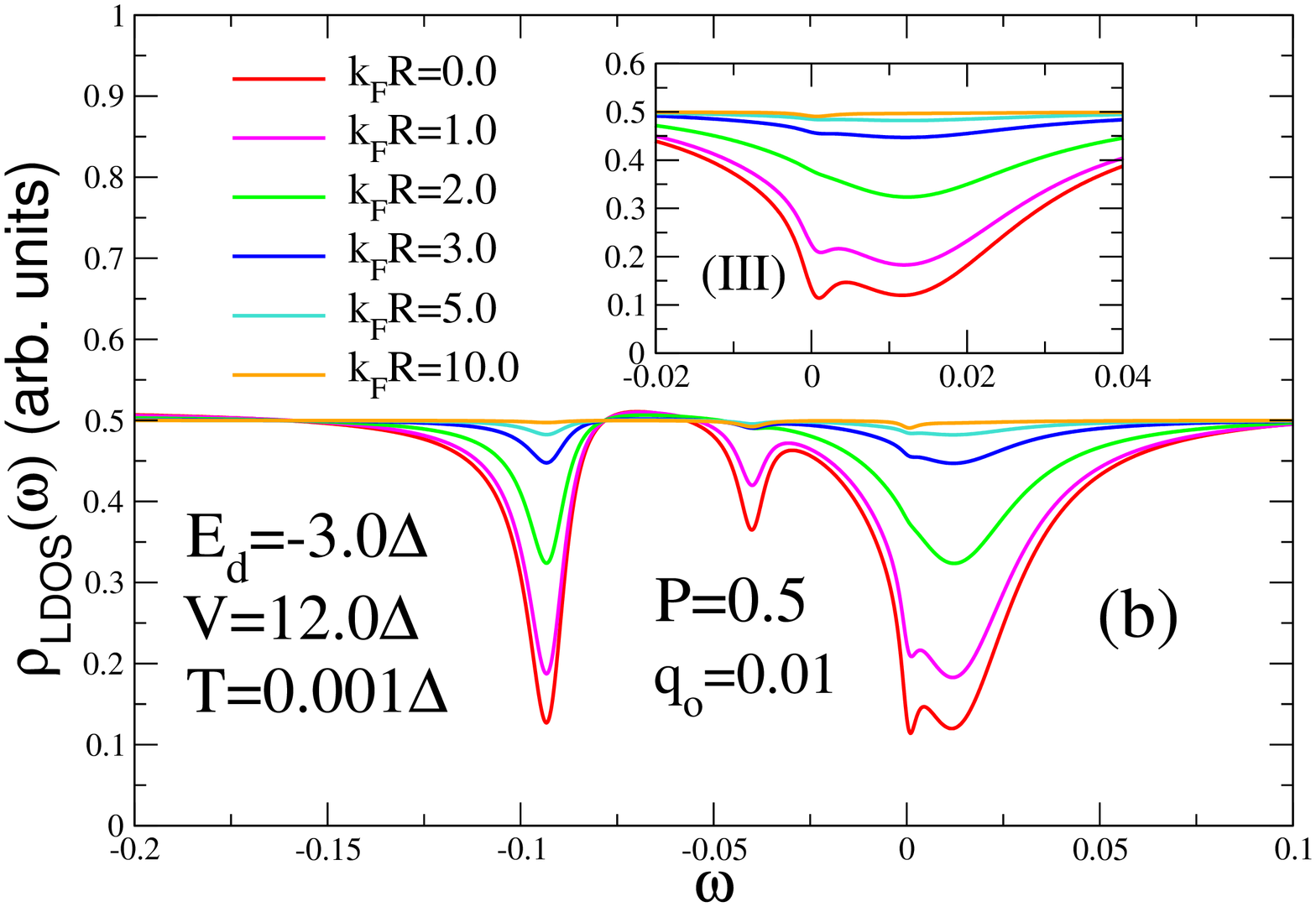}}}
\caption{ Spin-resolved local density of states (SR-LDOS) at $R=0$, in arbitrary units (a.u.),  as a function of the energy $\omega$.
In the full curve we represent the total density of states (LDOS). (a) In the inset (I), we present  the splitting of the Fano-Kondo
dip in the Kondo peak region for $E_{d}=-10\Delta$, whereas in the inset (II) we perform the same analysis for the intermediate valence case, using $E_{d}=-3\Delta$.
In the inset (III), we present the suppression of the double Fano-Kondo dip structure as a function of $k_{F}R$.}\label{kawaharashape}
\end{figure}

It is clear from Fig. \ref{kawaharashape}, that the resolved LDOS double dip structure is originated from the splitting of the
up and down spin components. To show the evolution of the double structure at $\omega\simeq\varepsilon_{F}=0$ from the Kondo peak to the
intermediate valence regime, we present in the inset (I) of Fig. (\ref{kawaharashape}) the splitting of the
Fano-Kondo dip for $E_{d}=-10\Delta$ whereas in the inset (II) we present the correspondent case in
the same energy range, but for the intermediate valence situation, with $E_{d}=-3\Delta$.
In this last regime, the spin down channel is also pinned at $\omega\simeq\varepsilon_{F}=0$, and the spin up is displaced
from it. Note that, for this set of parameters, the Fano-Kondo dip splitting is resolved in the LDOS. The  inset (II)
displays more precisely the spin-polarized antiresonances analogous to those observed in Fig. 1(b) of Ref. [\onlinecite{Kawahara2010}].
In Fig. \ref{kawaharashape}(b) we present the SR-LDOS as a function of the energy $\omega$ for different lateral STM-probe distances.
As $k_{F}R$ increases, the Fano-Kondo dips disappear gradually, and for $k_{F}R \simeq 10.0$, we recover the uncorrelated conduction band (see the inset (III)).
Similar behavior is observed in Fig. (3) of  Ref. [\onlinecite{Kawahara2010}].

\section{Conclusions}
\label{sec6}

We studied the spin-resolved local density of states (SR-LDOS) of a system composed of a Kondo adatom on a spin-polarized
two dimensional electron gas in the presence of an unpolarized STM probe. We derived the SR-LDOS expression in both the
presence and absence of the STM probe. Our expression to the SR-LDOS in terms of phase shifts is general
and independent of the method employed to calculate the local adatom GF. To determine this GF we
used the atomic approach in the limit of infinite $U$. The coupling parameter between adatom and host is assumed constant (local coupling).

We were able to study the SR-LDOS in all the interference regimes, varying the Fano factor $q_{o}$.
The main effect of the polarization was the tendency of one spin peak ($q_0=100$) or dip ($q_0=0.01$, $q_0=1$) in the SR-LDOS to remain pinned around
the host Fermi level as the polarization $P$ increases. In contrast, the other spin peak or dip is shifted and lose amplitude as $P$ increases.
This contrasts to the usual behavior in the presence of a magnetic field, where the Kondo resonance is symmetrically spin split
and destroyed as the magnetic field increases, while here it is enhanced in one channel (down) and destroyed in the other (up).

Our simulations are in close agreement with recent experimental results on adatom coupled
to a ferromagnetic island.\cite{Kawahara2010} The present system is a potential candidate
to promote the Kondo peak splitting without application of huge magnetic fields,
necessary for adatom systems characterized by a large $T_K$.
In particular, for the Fano factor $q_{o}=0.01$,
we observed a continuous second order insulator-metal transition driven by the polarization
as presented in Fig. \ref{dip1}(a). Finally,
our model was able to describe qualitatively several experimental results.\cite{SPSTM5,FM9,FM11,Kawahara2010,Nygard2004}

\begin{acknowledgments}
This work was supported by the Brazilian agencies CNPq, CAPES and FAPEMIG.
\end{acknowledgments}

\appendix

\section{The atomic GF for the Kondo adatom}

\label{ApA}

In this Appendix we present expressions employed in the spin dependent GF for the Kondo adatom considering the SIAM in the atomic version given by Eq. (\ref{SIAM2}) of Sec. \ref{secAP}. To obtain the atomic GF, we use Zubarev's notation \cite{Zubarev60}
\begin{eqnarray}
\mathcal{G}_{at,\sigma}^{dd}(\omega)& = &e^{\beta\Omega}\sum_{jj^{\prime}}\left(
e^{-\beta E_{j}}+e^{-\beta E_{j^{\prime}}}\right)\nonumber\\
& \times & \frac{|<j^{\prime}\;|\;X_{d,\sigma\sigma}\;|j>|^{2}}{\omega-(E_{j}%
-E_{j^{\prime}})}, \label{Zubarevo}%
\end{eqnarray}
where $\beta=1/k_{B}T$ with $k_{B}$ as the Boltzmann constant, $T$ is the system temperature and $\Omega$ is the thermodynamical potential. The eigenvalues $E_{j}$ and eigenvectors $|j>$ correspond to the complete solution of the SIAM Hamiltonian. The final result is the following
\begin{equation}
\label{GAnderson}\mathcal{G}_{at,\sigma}^{dd}(\omega) = e^{\beta\Omega} \sum_{i=1}^{8} \frac{m_{i\sigma}%
}{\omega-u_{i\sigma}},
\end{equation}
where the poles and the residues are
\begin{eqnarray}
u_{1\sigma} &=& E_{3\sigma}-E_{1\sigma} = E_{8\sigma}-E_{5\sigma} = E_{7\sigma}-E_{4\sigma}\nonumber\\
            &=& \frac{\scriptstyle 1}{\scriptstyle 2} \left(  \varepsilon_{dk\sigma} - \delta_{\sigma}\right)\nonumber\\
u_{2\sigma} &=& E_{5\sigma}-E_{1\sigma} = E_{8\sigma}-E_{3\sigma} = E_{7\sigma}-E_{2\sigma}\nonumber\\
            &=& \frac{\scriptstyle 1}{\scriptstyle 2} \left(  \varepsilon_{dk\sigma} + \delta_{\sigma}\right)\nonumber\\
u_{3\sigma} &=& E_{12\sigma}-E_{10\sigma} = \frac{\scriptstyle 1}{\scriptstyle 2} \left(  \varepsilon_{dk\sigma}- \delta_{\sigma}^{\prime}\right)\nonumber\\
u_{4\sigma} &=& E_{12\sigma}-E_{9\sigma} = \frac{\scriptstyle 1}{\scriptstyle 2} \left(  \varepsilon_{dk\sigma}+\delta_{\sigma}^{\prime}\right)\nonumber\\
u_{5\sigma} &=& E_{9\sigma}-E_{2\sigma} = \varepsilon_{k\sigma} - \frac{\scriptstyle 1}{\scriptstyle 2} \left(\delta_{\sigma}^{\prime}- \delta_{\sigma}\right)\nonumber\\
u_{6\sigma} &=& E_{10\sigma}-E_{2\sigma} = \varepsilon_{k\sigma} + \frac{\scriptstyle 1}{\scriptstyle 2} \left(\delta_{\sigma}^{\prime}+ \delta_{\sigma}\right)\nonumber\\
u_{7\sigma} &=& E_{9\sigma}-E_{4\sigma} = \varepsilon_{k\sigma} - \frac{\scriptstyle 1}{\scriptstyle 2} \left(\delta_{\sigma}^{\prime}+ \delta_{\sigma}\right)\nonumber\\
u_{8\sigma} &=& E_{10\sigma}-E_{4\sigma} =\varepsilon_{k\sigma} + \frac{\scriptstyle 1}{\scriptstyle 2} \left(\delta_{\sigma}^{\prime}- \delta_{\sigma}\right)\label{poles}
\end{eqnarray}
and
\begin{eqnarray}
m_{1\sigma} &=& c_{1\sigma}^{2}[ 1+ e^{-\frac{1}{2} \beta( \varepsilon_{dk\sigma}-\delta_{\sigma})}+\frac{\scriptstyle 3}{\scriptstyle 2} e^{-\frac{1}{2} \beta( \varepsilon_{dk\sigma}+\delta_{\sigma})}\nonumber\\
            &+& \frac{\scriptstyle 3}{\scriptstyle 2} e^{- \beta \varepsilon_{dk\sigma}} ]\nonumber\\
m_{2\sigma} &=& s_{1\sigma}^{2}[ 1+ e^{-\frac{ 1}{2} \beta(\varepsilon_{dk\sigma}+\delta_{\sigma})} + \frac{\scriptstyle 3}{\scriptstyle 2} e^{-\frac{1}{2} \beta(\varepsilon_{dk\sigma}-\delta_{\sigma})}\nonumber\\
            &+& \frac{\scriptstyle 3}{\scriptstyle 2} e^{- \beta \varepsilon_{dk\sigma}} ]\nonumber\\
m_{3\sigma} &=& c_{2\sigma}^{2}[e^{-\frac{ 1}{ 2} \beta( \epsilon_{d}+ 3\varepsilon_{k\sigma}+\delta_{\sigma}^{\prime})} +e^{-\frac{ 1}{ 2} \beta( \epsilon_{d}+ 2\varepsilon_{k\sigma})} ]\nonumber\\
m_{4\sigma} &=& s_{2\sigma}^{2}[ e^{-\frac{ 1}{ 2} \beta( \epsilon_{d}+ 3\varepsilon_{k\sigma}-\delta_{\sigma}^{\prime})} + e^{-\frac{ 1}{ 2} \beta( \epsilon_{d}+
2\varepsilon_{k\sigma})} ]\nonumber\\
m_{5\sigma} &=& \frac{\scriptstyle 1}{\scriptstyle 2}s_{1\sigma}^{2} c_{2\sigma}^{2}[ e^{-\frac{1}{2} \beta(\varepsilon_{dk\sigma}-\delta_{\sigma})} +e^{-\frac{1}{2} \beta(\epsilon_{d}+3\varepsilon_{k\sigma}-\delta_{\sigma}^{\prime})} ]\nonumber\\
m_{6\sigma} &=& \frac{\scriptstyle 1}{\scriptstyle 2} s_{1\sigma}^{2} s_{2\sigma}^{2}[e^{-\frac{1}{2} \beta(\varepsilon_{dk\sigma}-\delta_{\sigma})} + e^{-\frac{1}{2}
\beta(\epsilon_{d}+3\varepsilon_{k\sigma}+\delta_{\sigma}^{\prime})} ]\nonumber\\
m_{7\sigma} &=& \frac{\scriptstyle 1}{\scriptstyle 2} c_{1\sigma}^{2}\ c_{2\sigma}^{2}[e^{-\frac{1}{2} \beta(\varepsilon_{dk\sigma}+\delta_{\sigma})} + e^{-\frac{1}{2}
\beta(\epsilon_{d}+3\varepsilon_{k\sigma}-\delta_{\sigma}^{\prime})} ]\nonumber\\
m_{8\sigma} &=& \frac{c_{1\sigma}^{2} s_{2\sigma}^{2}}{2}[ e^{-\frac{\beta(\varepsilon_{dk\sigma}+\delta_{\sigma})}{2}}+
e^{-\frac{\beta(\epsilon_{d}+3\varepsilon_{k\sigma}+\delta_{\sigma}^{\prime})}{2}} ]\label{residues}
\end{eqnarray}
respectively, which are defined in terms of
\begin{eqnarray}
\epsilon_{d} &=& E_{d}-\epsilon_{F} \nonumber\\
\epsilon_{k\sigma} &=& E_{k\sigma}-\epsilon_{F} \nonumber\\
s_{1\sigma} &=& \sin\phi_{\sigma}\nonumber\\
c_{1\sigma} &=& \cos\phi_{\sigma}\nonumber\\
s_{2\sigma} &=& \sin\Lambda_{\sigma}\nonumber\\
c_{2\sigma} &=& \cos\Lambda_{\sigma}\nonumber\\
\epsilon_{d} &+& \varepsilon_{k\sigma}=\varepsilon_{dk\sigma},\nonumber\\
\delta_{\sigma} &=& ((\varepsilon_{k\sigma}-\epsilon_{d})^{2}+4V^{2})^{1/2}\nonumber\\
\delta_{\sigma}^{\prime} &=& ((\varepsilon_{k\sigma}-\epsilon_{d})^{2}+8V^{2})^{1/2}\label{definitions}
\end{eqnarray}
with $\epsilon_{F}$ being the Fermi energy and
\begin{eqnarray}
\tan\phi_{\sigma} &=& \frac{2V}{\varepsilon_{k\sigma}-\epsilon_{d}+\delta_{\sigma}}\nonumber\\
\tan\Lambda_{\sigma} &=& \frac{2\sqrt{2}V}{\varepsilon_{k\sigma}-\epsilon_{d}+\delta_{\sigma}^{\prime}}.\label{definitions2}
\end{eqnarray}

\section{Fano factor for the FM host}
\label{ApB}
In order to determine the Fano factor given by Eq. (\ref{eq:FM_Fano}) in Sec. \ref{sec3A} due to the
adatom-host coupling, we extend the procedure proposed for unpolarized bulk electrons \cite{GFM} to conduction states of a FM surface.
To that end we perform the calculation assuming the wide-band limit conditions $\omega\ll D_{\sigma}$
and $\Gamma\ll D_{\sigma}$ in the advanced GF
\begin{equation}
\tilde{G}_{\sigma}\left(\omega,R\right)=\frac{1}{N_{FM\sigma}}\sum_{\vec{k}}\frac{\Gamma^{2}}{\Gamma^{2}+\varepsilon_{k\sigma}^{2}}\frac{e^{i\vec{k}.\vec{R}}}{\omega-\varepsilon_{k\sigma}-i\eta} , \label{eq:ADV_GF}\end{equation}
with $\eta\rightarrow0^{+}$, which allows, in combination with Eq. (\ref{eq:Friedel}), to establish the following equalities
\begin{equation}
\Re\left\{ \tilde{G}_{\sigma}\left(\omega,R\right)\right\} =\Re\left\{ \tilde{g}_{\sigma}\left(\omega,R\right)\right\} \label{eq:real}\end{equation}
and
\begin{equation}
\Im\left\{ \tilde{G}_{\sigma}\left(\omega,R\right)\right\} =-\Im\left\{ \tilde{g}_{\sigma}\left(\omega,R\right)\right\} =\pi\rho_{0}A_{\sigma}\left(R\right) . \label{eq:imag-1}\end{equation}
Considering
\begin{equation}
J_{0}\left(kR\right)=\frac{1}{2\pi}\int_{0}^{2\pi}\exp\left[ikR\cos\theta_{k}\right]d\theta_{k}\label{eq:Jo}\end{equation}
as the angular representation for the zeroth-order Bessel function in Eq. (\ref{eq:Friedel}), and according to Eqs. (\ref{eq:FM_Fano}) and (\ref{eq:real}), the Fano factor becomes
\begin{equation}
q_{FM\sigma}=\frac{1}{\pi\rho_{0}}\Re\left\{ \tilde{G}_{\sigma}\left(\omega,R\right)\right\}.\label{eq:Fano_cb_}\end{equation}
We can calculate this parameter rewriting the equation above as

\begin{equation}
q_{FM\sigma}=\frac{1}{\pi\rho_{0}}\tilde{G}_{\sigma}\left(\omega,R\right)-iA_{\sigma}\left(R\right) , \label{eq:Fano_cb_II}\end{equation}
noting that the amplitude $A_{\sigma}\left(R\right)$ is already known from Eq. (\ref{eq:Friedel}).

Thus, the quantity $\frac{1}{\pi\rho_{0}}\tilde{G}_{\sigma}\left(\omega,R\right)$
must be found to provide the relationship for the Fano parameter, which can be done using the decomposition
\begin{equation}
\frac{1}{\pi\rho_{0}}\tilde{G}_{\sigma}\left(\omega,R\right)=\frac{1}{2}\frac{\rho_{FM\sigma}}{\rho_{0}}\sum_{l=1}^{2}\tilde{G}_{l}(\omega,R)  \label{eq:GI}\end{equation}
written in terms of the integral
\begin{eqnarray}
\tilde{G}_{l}(\omega,R) & =\frac{1}{\pi} & \int d\varepsilon_{k\sigma}H_{0}^{\left(l\right)}\left[k_{F\sigma}\left(1+\frac{\varepsilon_{k\sigma}}{D_{\sigma}}\right)R\right]\frac{\Gamma^{2}}{\Gamma^{2}+\varepsilon_{k\sigma}^{2}}\nonumber \\
 & \times & \frac{1}{\omega-\varepsilon_{k\sigma}-i\eta} , \label{eq:I_1}\end{eqnarray}
that depends on the Hankel functions $H_{0}^{\left(1\right)}(z)=J_{0}(z)+iY_{0}(z)$ and $H_{0}^{\left(2\right)}(z)=J_{0}(z)-iY_{0}(z)$. We conclude that the task here reduces to evaluate the integrals $\tilde{G}_{1}(\omega,R)$ and $\tilde{G}_{2}(\omega,R)$.

The first integral is calculated by choosing a counterclockwise contour over a semi-circle in the upper half of the complex plane that considers the simple pole $\varepsilon_{k\sigma}=+i\Gamma$. Following the residue theorem, we have
\begin{equation}
\tilde{G}_{1}(\omega,R)=H_{0}^{\left(1\right)}\left[k_{F\sigma}\left(1+i\frac{\Gamma}{D_{\sigma}}\right)R\right]\frac{\Gamma}{\omega-i\Gamma}.\label{eq:I1}\end{equation}
For the evaluation of the second integral we used a clockwise contour over a semi-circle in the lower half plane with poles placed at $\varepsilon_{k\sigma}=\omega-i\eta$ and $\varepsilon_{k\sigma}=-i\Gamma$, which leads to
\begin{eqnarray}
\tilde{G}_{2}(\omega,R)& = & 2iH_{0}^{\left(2\right)}\left[k_{F\sigma}\left(1+\frac{\omega}{D_{\sigma}}\right)R\right]\frac{\Gamma^{2}}{\Gamma^{2}+\omega^{2}}+\frac{\Gamma}{\omega+i\Gamma}\nonumber \\
 &  & \times H_{0}^{\left(2\right)}\left[k_{F\sigma}\left(1-i\frac{\Gamma}{D_{\sigma}}\right)R\right].\label{eq:I2}\end{eqnarray}
As the complex conjugate property $H_{0}^{\left(1\right)}\left(z^{*}\right)=\left[H_{0}^{\left(2\right)}\left(z\right)\right]^{*}$ is valid we are able to derive Eq. (\ref{eq:FM_Fano}) from Eq. (\ref{eq:Fano_cb_II}) considering Eqs. (\ref{eq:GI}), (\ref{eq:I1}) and (\ref{eq:I2}) in the wide-band limit characterized by the conditions $\omega\ll D_{\sigma}$ and $\Gamma\ll D_{\sigma}$.


\begin{thebibliography}{99}
\bibitem{comment1} Recently, though, it was found low Kondo temperatures (2.6 K) in a STM metallic system, comparable to typical $T_K$ in QDs (see Ref. [\onlinecite{STM12}]).
\end{thebibliography}
\end{document}